\RequirePackage{fix-cm}
\documentclass[smallextended]{svjour3}       % onecolumn (second format)
\smartqed  % flush right qed marks, e.g. at end of proof
\usepackage{graphicx}
\usepackage{mathptmx}      % use Times fonts if available on your TeX system
\usepackage{latexsym}
\usepackage{amsmath,amssymb,epsfig,rotating,graphicx,lscape,multirow}
\usepackage{natbib,amsmath,graphicx}
\DeclareMathOperator{\mean}{mean}
\DeclareMathOperator{\Var}{Var}
\journalname{Journal of Pharmacokinetics and Pharmacodynamics}
\begin{document}

\title{Cross-Validation for Nonlinear Mixed Effects Models}

\titlerunning{Cross-Validation for Nonlinear Mixed Effects Models}

\author{Emily Colby         \and
        Eric Bair %etc.
}


\institute{Emily Colby \at
              Dept. of Biostatistics \\
              Univ. of North Carolina-Chapel Hill \\
              Chapel Hill, NC \\
              \email{ecanimation@hotmail.com}           %  \\
           \and
           Eric Bair \at
              Depts. of Endodontics and Biostatistics \\
              Univ. of North Carolina-Chapel Hill \\
              Chapel Hill, NC \\
              \email{ebair@email.unc.edu}
}

\date{Received: Nov. 12, 2012 / Accepted: Mar. 10, 2013}

\maketitle

\begin{abstract}
Cross-validation is frequently used for model selection in a variety
of applications. However, it is difficult to apply cross-validation to
mixed effects models (including nonlinear mixed effects models or NLME
models) due to the fact that cross-validation requires
``out-of-sample'' predictions of the outcome variable, which cannot be
easily calculated when random effects are present. We describe two
novel variants of cross-validation that can be applied to nonlinear
mixed effects models. One variant, where out-of-sample predictions are
based on post hoc estimates of the random effects, can be used to
select the overall structural model. Another variant, where
cross-validation seeks to minimize the estimated random effects rather
than the estimated residuals, can be used to select covariates to
include in the model. We show that these methods produce accurate
results in a variety of simulated data sets and apply them to two
publicly available population pharmacokinetic data sets.

\end{abstract}

\section{Introduction}

\subsection{Overview of Population Pharmacokinetic and Pharmacodynamic Modeling}
Population pharmacokinetic and pharmacodynamic (PK/PD) modeling is the
characterization of the distribution of probable PK/PD outcomes
(parameters, concentrations, responses, etc) in a population of
interest. These models consist of fixed and random effects. The fixed
effects describe the relationship between explanatory variables (such
as age, body weight, or gender) and pharmacokinetic outcomes (such as
the concentration of a drug). The random effects quantify variation in
PK/PD outcomes from individual to individual.

Population PK/PD models are hierarchical. There is a model for the
individual, a model for the population, and a model for the residual
error. The individual PK model typically consists of a compartmental model of the
curve of drug concentrations over time. The pharmacokinetic
compartmental model is similar to a black box engineering model. Each 
of the compartments is like a black box, where a system of
differential equations is derived based on the law of conservation of
mass \citep{Sandler}. The number of such compartments to include in
the model must be determined based on the data.

The equations for the PK/PD parameters represent the model for the
population in the hierarchy of models. The PK/PD parameters are modeled
with regression equations containing fixed effects, covariates, and
random effects (denoted by $\eta$'s). The random effects account for
the variability across subjects in the parameters and for anything
left out of the parameter equations (such as a covariate not
included). The vector of random effects ($\eta$) is assumed to follow
a multivariate normal distribution with mean 0 and variance-covariance
matrix $\Omega$. The matrix $\Omega$ may be diagonal, full block, or
block diagonal. The model for the residual error ($\epsilon$) accounts
for any deviation from the model in the data not absorbed by the other
random effects. The residual error model may be specified such that
measurements with higher values are given less importance compared
with measurements with smaller values, often referred to as
``weighting''.

Hence, population PK/PD models are non-linear mixed effects (NLME)
models. They are represented by differential equations that may or may
not have closed-form solutions, and are solved either analytically or
numerically. The parameters are estimated using one of the various
algorithms available such as first order conditional estimation with
interaction (FOCEI). See \citet{wang} for a mathematical description
of these algorithms.

Once model parameters are estimated using an algorithm such as FOCEI,
one may fix the values of the model estimates and perform a post hoc
calculation to obtain random effect values ($\eta$'s) for each
subject. Thus, one may fit a model to a subset of the data and obtain
random effect values for the full data set. See \citet{wang} for a
discussion of how these posterior Bayes (post hoc) estimates of the
$\eta$'s are calculated.

\subsection{Cross-Validation and Nonlinear Mixed Effects Modeling}
In general, cross-validation is not frequently used for evaluating
nonlinear mixed effects (NLME) models \citep{eval}. When
cross-validation is applied to NLME models, it is generally used to
evaluate the predictive performance of a model that was selected using
other methods. For example, in \citet{propofol2}, data were pooled
across subjects to fit a model as though the data were obtained from a
single subject. Subjects were removed one at a time, and the accuracy
of the predicted observations with subsets of the data was
assessed. Another approach \citep{hookercv} removed subjects one at a
time to estimate model parameters and predicted PK parameters using
the covariate values for the subject removed. The parameters were
compared with the PK parameters obtained using the full data set in
order to evaluate the final model and identify influential
individuals. See \citet{mulla}, \citet{kerbusch}, and \citet{raja} for
additional examples where cross-validation was used to validate NLME
models.

Less frequently cross-validation is used for model selection in
NLME modeling. For example, one may wish to compare a model
with a covariate to another model without the covariate. In
\citet{ralph}, the prediction error of the posterior PK parameter for
each subject was calculated, and a paired t-test was performed to
compare the prediction error between a base and full model to assess
whether differences between the models were significant. The full
model was only found to be correct when the effect of the covariate
was large.

In several published studies, cross-validation failed
to identify covariate effects that were identified using other
methods. As noted earlier, \citet{ralph} found that cross-validation
only identified covariate effects when the effect was large.
Similarly, \citet{zomorodi} found that cross-validation tended to
favor a base model (without a covariate) despite the fact that the
covariate was found to be significant using alternative
approaches. \citet{fiset} also found that models with and without
covariates tended to produce comparable error rates despite the fact
that likelihood-based approaches favored models that included
covariates. Indeed, \citet{wahlby} used a special form of
cross-validation where one concentration data point was chosen for
each parameter, which was the point at which the parameter was most
sensitive based on partial derivatives. Once again, little difference
was observed between models that included covariates and corresponding
models without covariates. Thus, cross-validation can fail to detect
covariate effects even when attention is restricted to a subset of the
data that should be most sensitive to model misspecification.

Despite the fact that cross-validation may fail to detect covariate
effects, it has been successfully used to compare models with
structural differences, such as a parallel Michaelis-Menten and
first-order elimination (MM+FO) model and a Michaelis-Menten (MM)
model \citep{valodia}. This indicates that cross-validation can be
used for model selection in NLME modeling under certain
circumstances. Moreover, the fact that cross-validation often fails to
detect covariate effects is not surprising. When covariate effects are
present in an underlying NLME model, a misspecified model that fails
to include a covariate may not significantly decrease the predictive
accuracy of the model. This can occur when random effects in the
pharmacokinetic parameters can compensate for the missing
covariate. Thus, if cross-validation chooses the model with the lowest
out-of-sample prediction error, it may not be able to determine
whether a covariate should be included in the model.

Other methods have been proposed for using cross-validation for model
selection in NLME modeling \citep{ribbingposter, swcmv}. However,
these methods rely on estimation of the likelihood function, which is
unusual for cross-validation, and they have not been studied
extensively.

Thus, we propose an alternative form of cross-validation for covariate
model selection in NLME modeling. Rather than choosing a model which
minimizes the out-of-sample prediction error, we choose a model which
minimizes the post hoc estimates of the random effects ($\eta$'s). The
motivation is that if the $\eta$'s are large, this suggests that there
is a large amount of unexplained variation from individual to
individual, which indicates that a covariate may be missing from the
model. However, traditional cross-validation (which minimizes the
out-of-sample prediction error) is still useful for comparing
structural models, as we will discuss below.

\section{Methods}

\subsection{Cross-Validation}
Cross-validation is a method for evaluating the expected accuracy of a
predictive model. Suppose we have a response variable $Y$ and a
predictor variable $X$ and we seek to estimate $Y$ based on $X$. Using
the observed $X$'s and $Y$'s we may estimate a function $\hat{f}$ such
that our estimated value of $Y$ (which we call $\hat{Y}$) is equal to
$\hat{f}(X)$. Cross-validation is an estimate of the expected loss
function for estimating $Y$ based on $\hat{f}(X)$. If we use squared
error loss (as is conventional in NLME modeling), then
cross-validation is an estimate of
$E\left[\left(Y-\hat{f}(X)\right)^2\right]$.

A brief explanation of cross-validation is as follows: First, the data
is divided into $K$ partitions of roughly equal size. For the $k$th
partition, a model is fit to predict $Y$ based on $X$ using the $K-1$
other partitions of the data. (Note that the $k$th partition is not
used to fit the model.) Then the model is used to predict $Y$ based on
$X$ for the data in the $k$th partition. This process is repeated for
$k=1, 2, \ldots, K$, and the $K$ estimates of prediction error are
combined. Formally, let $\hat{f}^{-k}$ be the estimated value of $f$
when the $k$th partition is removed, and suppose the indices of the
observations in the $k$th partition are contained in $K_k$. Then the
cross-validation estimate of the expected prediction error is equal to
\[
\frac{1}{n} \sum_{i=1}^k \sum_{j \in K_i} \left(y_j -
  \hat{f}^{-i}(x_j) \right)^2
\]
Here $n$ denotes the number of observations in the data set. For a
more detailed discussion of cross-validation, see \citet{HTF08}. 

The above procedure is known as $k$-fold
cross-validation. Leave-one-out cross validation is a special case of
$k$-fold cross-validation where $k$ is equal to the number of
observations in the original data set.

\subsection{Comparing covariate models}
In some situations, a researcher may want to compare models with and
without covariate effects, such as a model with an age effect on
clearance versus a model without an age effect on clearance. This
method is designed to detect differences in models that affect the
equations for the parameters.

Consider a data set with subjects $i=1, 2, \ldots, n$.  Each subject
has observations $y_{ij}$ for $j= 1, 2, \ldots,t_{i}$ (where $t_{i}$
is the number of time points or discrete values of the independent
variable for which there are observations for subject $i$). The
question of interest is whether or not a fixed effect $dPdX$ for a
covariate $X$ should be included in an equation for a parameter $P$,
having fixed effect $tvP$ and random effect $\eta_{P}$. The equation
for $P$ could have any of the typical forms used in NLME modeling. For
example, one could compare a model with a covariate $X$
\begin{equation}
P = tvP*(X/\mean(X))^{dPdX}*\exp(\eta_{P})
\end{equation}
to a model having no covariate effect
\begin{equation}
P = tvP*\exp(\eta_{P})
\end{equation}
If a covariate $X$ has an effect on a parameter $P$, the unexplained
error in $P$ (modeled by $\eta_{P}$) when $X$ is left out of the model
tends to have higher variance. By including covariate $X$ in the
model, we wish to reduce the unexplained error in $P$, which is
represented by $\eta_{P}$. Therefore, metrics involving $\eta_{P}$ are
useful for determining whether a covariate $X$ is needed. Specifically,
one can perform cross-validation to compare the predicted $\eta_{P}$'s
when $X$ is included or not included in the model. We propose a
statistic for determining whether a covariate, $X$, is needed for
explaining variability in a parameter, $P$, when $P$ is modeled with a
random effect $\eta_{P}$. The statistic can be calculated as
follows:

For $i$ = 1 to $n$:
\begin{enumerate}
\item Remove subject $i$ from the data set.
\item Fit a mixed effects model to the subset of the data with subject
  $i$ removed.
\item Accept all parameter estimates from this model, and freeze the
  parameters to those values.
\item Fit the same model to the whole data set, without any major
  iterations, estimating only the post hoc values of the random
  effects. (In NONMEM, use the commands MAXITER=0, POSTHOC=Y. In NLME,
  set NITER to 0.)
\item Square the post hoc eta estimate for the subject that was left
  out for the parameter of interest
\end{enumerate}
Take the average of the quantity in step 5 over all subjects.

This sequence of steps can also be represented by the equation
\begin{equation}
\text{CrV}_\eta = \frac{1}{n} \sum\limits_{i=1}^{n}
  (\hat{\eta}_{P_{i,-i}})^{2}
\end{equation}
where $\hat{\eta}_{P_{i,-i}}$ is the post hoc estimate of the random
effect for the $i$th subject for parameter $P$ in a model where the
$i$th subject was removed, and $n$ is the number of subjects. Note
that our method leaves out one subject at a time, rather than one
observation at a time.

In general, one will favor the model with the minimum value of
$\text{CrV}_\eta$. However, to avoid over-fitting, it is common when
applying cross-validation to choose the most parsimonious model
(i.e. the model with the fewest covariates) that is within one
standard error (SE) of the model with minimum $\text{CrV}_\eta$
\citep{HTF08}. We will follow this convention in all of our subsequent
examples. We define $\text{SE}(\text{CrV}_\eta)$ as the sample standard
deviation of the squared post hoc etas for the subjects left out divided
by the square root of the number of subjects. The formula for
$\text{SE}(\text{CrV}_\eta)$ is given by

\begin{equation}
\text{SE}(\text{CrV}_\eta) =
\sqrt{\frac{1}{n(n-1)}\sum\limits_{i=1}^{n}
  (x_{i} - \bar{x_{i}})^{2}}
\end{equation}
where 
\begin{equation}
x_{i} = \hat{\eta}_{P_{i,-i}}^{2}
\end{equation}

Alternatively, one may follow the same procedure while removing more
than one subject at a time. For example, one may divide the data into
$k$ roughly equally-sized partitions, fit a model using the data in
$k-1$ of the partitions, and calculate the post hoc $\eta$ values for
the subjects left out of the model. For data sets with large numbers
of subjects, this approach is obviously faster than the
``leave-one-out'' approach, and it may also reduce the amount of
variance in the cross-validation estimates \citep{HTF08}. However,
this approach may not be practical if the number of subjects is
small. We will only consider the leave-one-out method in our
subsequent analysis.

\subsection{Comparing models with major structural differences}
In other situations, a researcher may want to compare models with
major structural differences, such as a one compartment model
and a two compartment model. This method is designed to detect
differences in models that affect the overall shape of the response.

As discussed previously, consider a data set with subjects $i=1, 2,
\ldots, n$, where each subject has observations $y_{ij}$ for $j=1, 2,
\ldots ,t_{i}$. The statistic can be calculated as follows:

For $i$ = 1 to $n$:
\begin{enumerate}
\item Remove subject $i$ from the data set.
\item Fit a mixed effects model to the subset of the data with subject
  $i$ removed.
\item Accept all parameter estimates from this model, and freeze the
  parameters to those values.
\item Fit the same model to the whole data set, without any major
  iterations, estimating only the post hoc values of the random
  effects. (In NONMEM, use the commands MAXITER=0, POSTHOC=Y. In NLME,
  set NITER to 0.)
\item Calculate predicted values for subject $i$ (the subject that was
  left out). Note that this estimate uses the post hoc estimate of the
  random effects for subject $i$.
\item Take the average of the squared individual residuals for the
  subject that was left out (over all time points or over all values
  of the independent variable $t_{i}$)
\end{enumerate}
Take the average of the quantity in step 6 over all subjects.

This sequence of steps can also be represented by the equation
\begin{equation}
\text{CrV}_y = \frac{1}{n} \sum\limits_{i=1}^{n}\frac{
    \sum\limits_{j=1}^{t_i}
    (y_{ij} - \hat{y}_{ij,-i})^{2}}
    {t_i}
\end{equation}
where $y_{ij}$ is the observed value for the $i$th subject at the
$j$th time point or independent variable value and $\hat{y}_{ij,-i}$
is the predicted value for the $i$th subject at the $j$th time point
or independent variable value in a model where subject $i$ is left out
and post hocs are obtained. Once again, note that our method leaves
out one subject at a time, rather than one observation at a time.

For purposes of exploration, another statistic that takes into account
the weighting of the response can also be calculated:
\begin{equation}
\text{wtCrV}_y = \frac{1}{n} \sum\limits_{i=1}^{n}\frac{
    \sum\limits_{j=1}^{t_i}
    \text{WTIRES}_{ij,-i}^{2}}
    {t_i}
\end{equation}
Here $\text{WTIRES}_{ij,-i}$ is the individual weighted residual for
subject $i$ at time or independent variable value $j$ in a model where
subject $i$ is left out and post hocs are obtained, which is defined
to be:
\begin{equation}
  \text{WTIRES}_{ij,-i} = \frac{\sqrt{wt_{ij,-i}}(y_{ij} -
    \hat{y}_{ij,-i})}{\hat{\sigma}_{-i}}
\end{equation}
where $wt_{ij,-i}$ is the weight defined by the residual error model
(equal to the squared reciprocal of $\hat{y}_{ij,-i}$ for constant CV
error models or 1 for additive error models), and
$\hat{\sigma}_{-i}^{2}$ is the estimated residual variance.

As discussed previously, we will follow the convention of choosing the
most parsimonious model (defined to the model with the fewest number
of compartments) within one standard error (SE) of the model with the
minimum $\text{CrV}_y$. The formula for $\text{SE}(\text{CrV}_y)$ is
given by
\begin{equation}
\text{SE}(\text{CrV}_y) = \sqrt{\frac{1}{n(n-1)}\sum\limits_{i=1}^{n}
  (x_{i} - \bar{x_{i}})^{2}}
\end{equation}
where 
\begin{equation}
x_{i} = \frac{\sum\limits_{j=1}^{t_i}
      (y_{ij} - \hat{y}_{ij,-i})^{2}}
      {t_i}
\end{equation}
and the formula for $\text{SE}(\text{wtCrV}_y)$ is calculated
similarly, with 
\begin{equation}
x_{i} = \frac{
    \sum\limits_{j=1}^{t_i}
    \text{WTIRES}_{ij,-i}^{2}}
    {t_i}
\end{equation}
One may also consider $k$-fold cross-validation, although we will
restrict our attention to leave-one-out for our subsequent analysis.

This method is similar to existing methods for cross-validation on
NLME models. However, some applications of cross-validation do not use
post hoc estimates of the outcome variable, which is an important
difference from our proposed method. Also, we will show why this
method should not be used for comparing covariate models.

\subsection{Simulated Data Analysis}
Five sets of simulated data were generated to evaluate the performance
of our proposed cross-validation methods. In each simulation scenario,
two models were compared: a sparser ``base model'' and a less
parsimonious ``full model.'' The objective was to determine which of
the two possible models was correct using cross-validation.

A brief description of the five simulation scenarios is given in Table
\ref{T:sim_desc}. For a more detailed description of how the simulated
data sets were calculated, see Section \ref{SS:sim_details} in Online
Resource 1. For simulation scenarios 1-4, 200 simulated data sets
were generated using Pharsight's Trial Simulator software version
2.2.1. (Only 100 simulated data sets were generated for scenario 5.)
For each simulated data set, Pharsight's Phoenix NLME (platform
version 1.3) was used to fit the appropriate population PK models
(both the base model and the full model) using the Lindstrom-Bates 
method \citep{LB}. The $\eta$ shrinkage of each model was calculated
and diagnostics were performed to verify the convergence of each
model. To calculate the cross-validation statistics for each simulated
data set, subjects were removed from the data set one at a time and
the models were recalculated with each subject removed. Post hoc
estimates of the random effects (and corresponding predicted values of
$y$) were then calculated for the subject that was excluded from the
model. The values of $\text{CrV}_\eta$, $\text{CrV}_y$, and
$\text{wtCrV}_y$ were obtained by averaging over each such model. The
simulated data sets, batch files, Phoenix mdl files, and other files
used to process the output are available from the authors by request.

\begin{table}
\caption{Description of the five simulation scenarios}
\label{T:sim_desc}
\begin{tabular}{lp{3cm}p{3cm}p{3cm}}
\hline\noalign{\smallskip}
Scen. & True Model & Base Model & Full Model  \\
\noalign{\smallskip}\hline\noalign{\smallskip}
1 & one compartment, no covariate effects & one compartment, no
covariate effects & one compartment, age effect on clearance \\
2 & one compartment, age effect on clearance & one compartment, no
covariate effects & one compartment, age effect on clearance \\
3 & two compartments, age effect on clearance & two compartments, no
covariate effects & two compartments, age effect on clearance \\
4 & one compartment, body weight effect on volume, body weight, age,
gender, and hepatic impairment effects on clearance & one compartment,
body weight effect on volume, body weight, age, and gender effects on
clearance & one compartment, body weight effect on volume, body
weight, age, gender, and hepatic impairment effects on clearance \\
5 & two compartments, no covariate effects & one compartment, no
covariate effects & two compartments, no covariate effects \\
\noalign{\smallskip}\hline
\end{tabular}
\end{table}

For each simulated data set, the base model was selected if the value
of $\text{CrV}_\eta$ for the base model was less than that of the full
model. The full model was selected if the value of $\text{CrV}_\eta$
was less than that of the base model plus one standard error (using
the convention that the more parsimonious model is preferable if its
cross-validation error is within one standard error of the
cross-validation error of a less parsimonious model). Similar decision
rules were used for $\text{CrV}_y$ and $\text{wtCrV}_y$. The Akaike's
information criterion (AIC) \citep{hA74} and Bayesian information
criterion (BIC) \citep{gS78} were also calculated for the two models
for each simulated data set. The model (base or full) with the smallest
AIC/BIC was selected under the two criteria. For each scenario, the
performance of cross-validation was compared to the performance of
AIC/BIC using a two-sample proportion test.

Note: Consistent with the recommendations of \citet{VC97}, the BIC was
weighted by the number of observations. Although others have suggested
that the BIC should be weighted by the number of subjects\citep{KR95},
one recent simulation study found that neither choice of weight
consistently outperforms the other when applied to mixed models
\citep{mG06}.

\subsection{Indomethacin Data Analysis}
Pharsight’s Phoenix NLME (version 1.3) was used to fit models to
a previously published indomethacin data set \citep{indometh}.  The
data consists of six subjects with 11 observations per subject. Each
subject was administered a 25 mg dose of indomethacin intravenously at
the beginning of the study, and the concentration of indomethacin was
measured at 11 time points over an eight-hour period.

\begin{figure}
\includegraphics[scale=1]{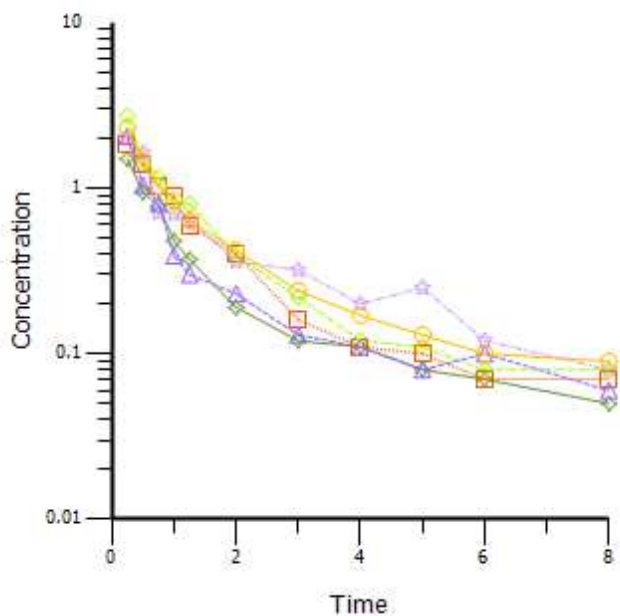}
\caption{Concentration versus time profiles from the indomethacin
  data set}
\label{F:indometh}
\end{figure}

The concentrations were plotted versus time for each subject (see
Figure \ref{F:indometh}). Based on the plot, a two compartment IV
bolus model with clearance parametrization and a proportional residual
error model was fit to this data set. See Section
\ref{SS:indometh_desc} in Online Resource 1 for a more detailed
description of the model. Individual initial estimates were obtained
using the curve stripping method \citep{Gibaldi} with a WinNonlin
Classic model. The averages of the individual PK parameters were used
as initial estimates for the pop PK model. Random effects were added
to the PK parameters for systemic volume and clearance in the form
$\theta P*\exp(\eta P)$, where $P$ is the parameter of interest. The
Phoenix project file used to fit this model is available from the
authors by request.

After fitting the model, a series of diagnostic plots were generated
to assess the validity of the model. (See Section
\ref{SS:indometh_diag} in Online Resource 1 for details.) The model
was then compared to a one compartment model based on both
$\text{CrV}_y$ and a likelihood ratio test (LRT). First, the model was
refit without including any random effects on the PK parameters. (The
LRT cannot be used when the random effects are included in this case
since the one compartment model forces the removal of some random
effects, which implies that these random effects have variances of
0. Thus, comparing the two models would require testing the null
hypothesis that a variance is equal to 0, which is on the boundary of
the parameter space, rendering the LRT invalid. See \citet{FLW11} for
details.). The LRT was used to test the null hypothesis of no
difference in the predictive accuracy of the two compartment model
versus the one compartment model. The value of $\text{CrV}_y$ was also
calculated for both models. Finally, the value of $\text{CrV}_y$ was
calculated for a one compartment and two compartment version of the
original model (with random effects included). See Section
\ref{SS:indometh_desc} in Online Resource 1 for a detailed description
of the models that were considered.

\subsection{Theophylline Data Analysis}
Pharsight’s Phoenix NLME (version 1.3) was used to fit models to a
published theophylline data set \citep{theo}. This theophylline
data set consists of twelve subjects with eleven observations per
subject. Each subject was administered a dose of theophylline at the
beginning of the study ranging between 3.1 mg/kg and 5.86 mg/kg. The
concentration of theophylline was measured at 11 time points per
subject over a 24 hour period. Each subject's weight was also
recorded.

\begin{figure}[here]
\includegraphics[scale=1]{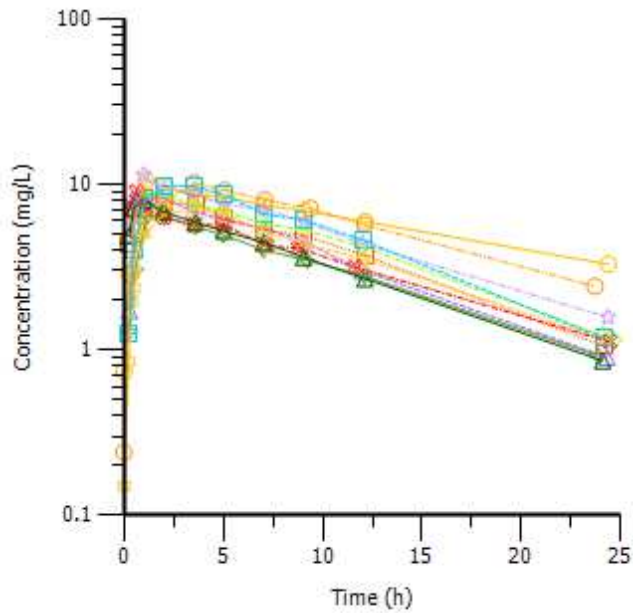}
\caption{Concentration versus time profiles from the theophylline
  data set}
\label{F:theo}
\end{figure}

The concentration were plotted versus time for each subject (see
Figure \ref{F:theo}). Based on the plot, a one compartment
extravascular model with clearance parametrization and an additive
residual error model was fit to this data set. Random effects were
added to the PK parameters for absorption rate, and systemic volume
and clearance in the form $\theta P*\exp(\eta P)$, where $P$ is the
parameter of interest. See Section \ref{SS:indometh_desc} in Online
Resource 1 for a more detailed description of the model. The
Phoenix project file used to fit this model is available from the
authors by request.

\begin{figure}[here]
\includegraphics[scale=0.9]{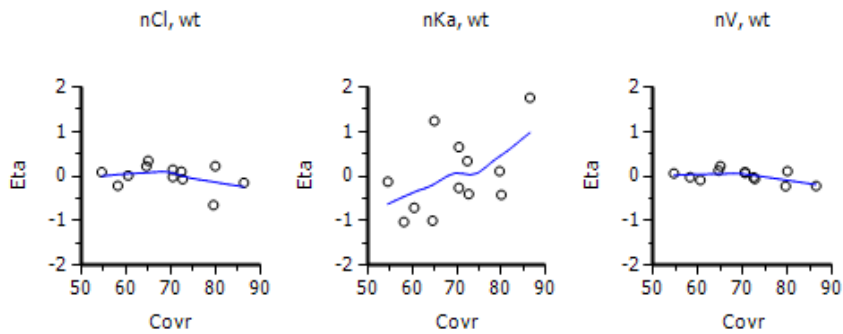}
\caption{$\eta$ versus covariate plots for the theophylline model with
  Tlag and no weight effect on Ka}
\label{F:theo2}
\end{figure}

The LRT and the $\text{CrV}_y$ statistic were used to compare a model
with a time lag parameter (Tlag) to a model without a Tlag parameter,
(with no random effect on the Tlag parameter). Moreover, the covariate
plots for the model with Tlag seemed to indicate a body weight effect
on $Ka$ might be needed (see Figure \ref{F:theo2}). Thus, the LRT and
the $\text{CrV}_\eta$ statistic were used to compare a model with the
Tlag parameter and body weight effect on $Ka$ to the model with the
Tlag parameter and no body weight covariate.

\section{Results}

\subsection{Simulation Results}
The results of the simulations are summarized in Table
\ref{T:sim_correct}. The $\text{CrV}_\eta$ statistic was correct in
97.0 percent of the 200  cases under scenario 1, whereas AIC was
correct in 88.5 percent of cases and BIC was correct in 94.5 percent
of cases. It correctly identified the full model under scenario 2 in
92.5 percent of the 200 cases, whereas AIC found the correct model in
98.5 percent of cases and BIC found the correct model in 93 percent of
cases. Under scenario 3, $\text{CrV}_\eta$ was correct in 93.0 percent
of cases, whereas AIC and BIC were correct in 97.5 and 94.0 percent of
cases, respectively. Under scenario 4, $\text{CrV}_\eta$ was correct
in 97.0 percent of cases, whereas AIC and BIC were correct in 71.5 and
64.0 percent of cases, respectively. $\text{CrV}_\eta$ was
significantly more likely to identify the correct model than AIC in
scenario 1 ($p=0.002$) and it was significantly more likely to
identify the correct model than both AIC and BIC in scenario 4
($p<0.0001$ in both cases). The performance of $\text{CrV}_y$ and
$\text{wtCrV}_y$ was also evaluated for scenarios 1-4, but it
performed poorly in each case with the exception of scenario 1. All
four applicable methods (AIC, BIC, $\text{CrV}_y$, and
$\text{wtCrV}_y$) correctly identified the true model under scenario 5
in 100 out of 100 cases.

\begin{table}
\caption{Proportion of times model comparison methods were correct out
  of 200 replicates (and associated standard error)}
\label{T:sim_correct}
\begin{center}
\begin{tabular}{ccccccc}
\hline\noalign{\smallskip}
True Model& Comparison& AIC& BIC& $\text{wtCrV}_y$& $\text{CrV}_y$&
$\text{CrV}_\eta$ \\
\noalign{\smallskip}\hline\noalign{\smallskip}
1 Cpt & 1 Cpt, Age-Cl&  0.885 (0.023)&	0.945 (0.016)&	0.940 (0.017)&
0.965 (0.013)& 0.970 (0.012)\\
1 Cpt, Age-Cl & 1 Cpt&	0.985 (0.009)& 0.930 (0.018)& 0.0 (0)& 0.0 (0)& 0.925 (0.018)\\
2 Cpt, Age-Cl & 2 Cpt&	0.975 (0.011)& 0.940 (0.017)& 0.005 (0.005)&
0.010 (0.007)& 0.930 (0.018)\\
1 Cpt, BW-V; &  1 Cpt, BW-V; & 0.715 (0.032)& 0.640 (0.034)& 0.015 (0.009)& 0.005 (0.0005) & 0.970 (0.012)\\
BW-Cl, G-Cl, & BW-Cl, G-Cl, & & & & & \\ 
Age-Cl, HI-Cl & Age-Cl & & & & & \\ 
2 Cpt & 1 Cpt &	1.0 (0)*&	1.0 (0)*& 1.0 (0)*&  1.0 (0)*& N/A\\
\noalign{\smallskip}\hline
\end{tabular}
\end{center}
Cpt=Compartment, Age-Cl indicates age effect on clearance, BW=Body
Weight, V=Volume, G=Gender, HI=Hepatic Impairment \\
*Based on 100 replicates
\end{table}

Some additional information about the distributions of the various test
selection statistics are contained in Tables \ref{T:sim_AICBIC},
\ref{T:sim_npress}, \ref{T:sim_mpress}, and \ref{T:sim_wtmpress} in
Section \ref{SS:sim_details} in Online Resource 1. In general the mean
values of AIC and BIC are lower in the true models (compared to the
misspecified values) and the mean value of $\text{CrV}_\eta$ is
significantly lower in the true models. This is not true for
$\text{CrV}_y$ and $\text{wtCrV}_y$ in scenarios 1-4; both statistics
tend to be smaller for the base model irrespective of which model is
correct (which explains the poor performance of these statistics in
scenarios 2-4). It is also noteworthy that an outlying observation
generated an extreme value for $\text{wtCrV}_y$ for one simulated data
set in scenario 3.

Boxplots of the $\eta$ shrinkage values for both models under
scenarios 1-4 are shown in Figure \ref{F:shrinkage} in Online Resource
1. (The $\eta$ shrinkage values were not calculated for scenario 5
since $\text{CrV}_\eta$ was not used in this scenario.) The models
converged for all simulated data sets with the exception of two
instances of scenario 5 (although some instances of all five simulated
scenarios produced models that showed signs of numerical
instability).

\subsection{Indomethacin Example} 
The final model appeared to fit well based on the diagnostic plots
(see Figures \ref{F:indometh2}, \ref{F:indometh3}, \ref{F:indometh4}
in Online Resource 1). The model coefficients and shrinkage estimates
are shown in Tables \ref{T:theta_indometh} and \ref{T:omega_indometh}
in Online Resource 1.

The LRT favored the two compartment model (with no random effects)
over the corresponding one compartment model ($p<0.0001$). The
$\text{CrV}_y$ statistic was in agreement with the LRT, having a value
of 0.1419 (SE 0.03393) for the one compartment model, and 0.0428 (SE
0.01355) for the two  compartment model. The $\text{CrV}_y$ statistic
in the model with random effects also favored the full (two
compartment) model over the base (one compartment) model. The value of
$\text{CrV}_y$ for the full model was 0.01679 (SE 0.004194) and 0.1406
(SE 0.03358) for the base model.

\subsection{Theophylline Example}
The final model appeared to fit well based on the diagnostic plots
(see Figures \ref{F:theoph2a}, \ref{F:theoph3}, \ref{F:theoph4}
in Online Resource 1). The model coefficients and shrinkage estimates
are shown in Tables \ref{T:theta_theoph} and \ref{T:omega_theoph} in
Online Resource 1.

The LRT favored the Tlag model ($p<0.0001$). The $\text{CrV}_y$
statistic was in agreement with the LRT, having a value of 0.2546 (SE
0.05727) for the model with Tlag, and 0.3927 (SE 0.10001) for the
model without Tlag. The LRT had a borderline result ($p=0.0667$) for
comparing the model with a body weight effect on $Ka$ and Tlag to the
model without a body weight effect on $Ka$ and Tlag, while the $\eta$
versus covariate plot (Fig \ref{F:theo2}) indicated an effect. The
$\text{CrV}_\eta$ statistic clearly favored the full model with a Tlag
and a weight effect on $Ka$, having a value of 0.06220 (SE 0.02942)
for the full (Tlag and wt) model, and 0.7819 (SE 0.2846) for the base
(Tlag) model.

\section{Discussion}
As noted earlier, cross-validation is not frequently used for
comparing NLME models \citep{eval}. Other methods, such as
the LRT, AIC, and BIC are more commonly used. However, each of these
alternative approaches have certain drawbacks. All three methods can
only be applied to models having the same residual error model. The
LRT can only be applied when models are nested and both models have
the same random effects. Moreover, there may be an inflated type I
error rate associated with the LRTs \citep{julie}. Both the AIC and
BIC have other shortcomings as well. Specifically, the AIC tends to
overfit, meaning that it keeps too many covariates in the model. The
BIC, in contrast, tends to underfit (fail to include significant
covariates), particularly when the same size is small \citep{HTF08}.

Our results show that cross-validation can be used for model selection
in NLME modeling and that it can produce significantly better results
than these competing methods. The $\text{CrV}_\eta$ statistic
identified the correct model at least 92.5\% of the time in each of
the simulated examples we considered. In contrast, the AIC was
significantly more likely to select a covariate for age in our first
simulation scenario (even though age had no effect on clearance in the
simulated model), and both the AIC and BIC were significantly less
likely to detect the effect of hepatic impairment in our fourth
simulation scenario.

All four applicable methods (AIC, BIC, $\text{CrV}_y$, and
$\text{wtCrV}_y$) correctly identified the true (two compartment) in
our fifth simulation scenario in 100 out of 100 cases. This finding is
of interest because the standard likelihood ratio test cannot be
applied when there are random effects in the full model that 
are not present in the base model.

Both the $\text{CrV}_y$ statistic and the LRT favored a two
compartment model in the indomethacin example. However, in the
theophylline example, the population covariate plots ($\eta$'s versus
covariates) seemed to suggest a weight covariate should be included in
the model even though the LRT was not significant at the $p<0.05$
level. The $\text{CrV}_\eta$ statistic clearly favored the model with
the weight covariate. This is consistent with previous studies showing
that theophylline distributes poorly into body fat. Hence, the
administered mg/kg dose should be calculated on the basis of ideal
body weight, \citep{pG78, tR82} implying that body weight affects the
extent of absorption. This is possible evidence that the LRT cannot
always identify covariate effects when they exist and that
cross-validation may be able to detect covariate effects in these
situations.

Although it may seem reasonable to use the $\text{CrV}_y$ or
$\text{wtCrV}_y$ statistics to determine if a covariate should be
included in a model, our simulations suggest that they can give
misleading results. Both statistics consistently favored models
without a covariate even when a covariate effect existed. The
predicted values are just as accurate with and without the covariate
effect when the true model has a covariate effect, because the
$\eta$'s can compensate for a missing covariate in a parameter. This
suggests that one should use $\text{CrV}_\eta$ rather than
$\text{CrV}_y$ or $\text{wtCrV}_y$ in situations when one wishes to
compare different covariate models. This also indicates that it may be
misleading to use cross-validation for model validation (as opposed to
model selection) if one uses post hoc estimates of the $\eta$'s when
calculating the predicted value of the response on the ``left out''
portion of the data. One may obtain a low cross-validation error rate
even when the model is misspecified.

One possible drawback to using cross-validation for model selection is
the fact that it is more computationally intensive than the LRT, AIC,
or BIC. Leave-one-out cross-validation was applied in each of the
examples in the present study, since each example consisted of
relatively small data sets. However, larger data sets may require 1-2
hours (or as many as 10 hours in extreme cases), to fit a single
model. If such a data set included hundreds of subjects,
leave-one-out cross-validation would clearly be computationally
intractible. In such situations, one may reduce the computing time by
reducing the number of cross-validation folds. If 10-fold
cross-validation is performed, this requires that the model be fitted
only 10 times, and the number of folds could be reduced further if
needed. Even a complicated model that required ten hours to fit could
be evaluated over the course of several days using 5-fold
cross-validation. Indeed, these cross-validation methods are no more
computationally intensive than bootstrapping, which is commonly used
to validate NLME models. The extra computational cost may be
worthwhile in situations where it is important that the model is
specified correctly.

Another potential issue with cross-validation is the fact that
estimation methods for NLME models sometimes fail to converge.
Although this was not a major issue in the examples we considered,
if the model fails to converge for a significant proportion of the
cross-validation folds, it is possible that it will produce inaccurate
results. Further research is needed on the effects of lack of
convergence on our proposed cross-validation methods.

These methods might be applied more generally with modifications to
other types of linear mixed effects models or generalized linear mixed
effects models. These methods may be applied without modification to
population PK/PD models and sparser data. These are areas for future
research. We expect in the sparse data case that the effectiveness of
the covariate selection method may be compromised by $\eta$ shrinkage,
which could distort the $\eta$ size criterion. The covariate selection
method introduced in this paper may not produce correct results for
parameters with high $\eta$ shrinkage (greater than 0.3, for example,
in a model where the covariate is not included). The random effects
for those parameters are typically removed during model development,
and hence covariate adjustments may not be needed for those
parameters. However, cross-validation produced correct results in our
first simulation scenario even though the median shrinkage was
approximately 0.3 in the base model (Figure \ref{F:shrinkage} in
Online Resource 1). The conditions under which our proposed
cross-validation method produces valid results in sparse data sets is
another area for future research.

\begin{acknowledgements}
Eric Bair was partially supported by National Institute of
Environmental Health Sciences/National Institutes of Health grant
P30ES010126 and National Center for Advancing Translational
Sciences/National Institutes of Health grant UL1RR025747. Emily Colby
was employed at Pharsight while preparing this manuscript. Bob Leary
and Dan Weiner provided valuable input. Tom Colby helped with
programming.
\end{acknowledgements}

\bibliographystyle{spbasic}      % basic style, author-year citations
\bibliography{bibliography}   % name your BibTeX data base

\newpage

\setcounter{page}{1}

\noindent
Article Title: Cross-Validation for Nonlinear Mixed Effects Models\\
Journal Name: Journal of Pharmacokinetics and Pharmacodynamics\\

\noindent
Emily Colby\\
Pharsight - A Certara Company\\
Cary, NC\\
and Dept. of Biostatistics\\
Univ. of North Carolina-Chapel Hill\\
Chapel Hill, NC\\
emily.colby@certara.com\\

\noindent
Eric Bair\\
Depts. of Endodontics and Biostatistics\\
Univ. of North Carolina-Chapel Hill\\
Chapel Hill, NC\\
ebair@email.unc.edu

\newpage

\renewcommand{\thesection}{S\arabic{section}}
\setcounter{section}{0}
\renewcommand{\thefigure}{S\arabic{figure}}
\setcounter{figure}{0}
\renewcommand{\thetable}{S\arabic{table}}
\setcounter{table}{0}

\section{Simulation Details} \label{SS:sim_details}

\subsection{Notation} \label{SS:notation}
In the subsequent examples, we denote the following population PK
parameters with the following symbols:
\begin{itemize}
\item $C$: concentration of the drug in central compartment
\item $C2$: concentration of the drug in peripheral compartment
\item $\text{CObs}$: observed concentration of the drug (which is
  measured with error)
\item $C_\epsilon$: error associated with $\text{CObs}$
\item $Aa$: amount of the drug in the absorption compartment
\item $A1$: amount of the drug in the central compartment
\item $A2$: amount of the drug in the peripheral compartment
\item $Ka$: absorption rate parameter
\item $V$: systemic volume parameter
\item $V2$: volume of peripheral compartment parameter
\item $Cl$: systemic clearance parameter
\item $Cl2$: clearance of distribution parameter
\end{itemize}
When any of the parameters above is preceded by $tv$ (i.e. $tvKa$),
this represents a fixed effect (which is fixed for a given simulation
but may vary across the replicates of the simulated data sets), and
when a variable is preceded by $\eta$ (i.e. $\eta Ka$), this
represents a random effect (which varies from subject to subject
within a given simulated data set).

\subsection{Simulation Example 1: No Covariate Effects}
A one-compartment, extravascular model was simulated with eight
subjects using Pharsight's Trial Simulator software (version
2.2.1). The equations for the model are as follows:
\begin{align*}
\frac{dAa}{dt} &= - Ka \cdot Aa \\
\frac{dA1}{dt} &= Ka \cdot Aa - Cl \cdot C \\
C &= \frac{A1}{V} \\
\end{align*}
A 10 percent constant CV percentage was simulated for the residual
error:
\[
\text{CObs} = C(1 + C_\epsilon)
\]
where $\Var(C_\epsilon)=0.01$.

The absorption rate parameter, $Ka$, was simulated with only a fixed
effect. All other parameters were simulated with fixed and random
effects as follows:
\[
\begin{array}{rl}
    Ka &= tvKa \\
    V &= tvV \cdot \exp(\eta V) \\
    Cl &= tvCl \cdot \exp(\eta Cl) \\
\end{array}
\]
The fixed effects for the PK parameters were assumed to be normally
distributed at the study level (varying across replicates) with means
listed below and standard deviations of 0.1:
\[
\begin{array}{rl}
    \mean(tvKa) &= 0.35 \\
    \mean(tvV) &= 13.5 \\
    \mean(tvCl) &= 7.4 \\
\end{array}
\]

The random effects ($\eta V$ and $\eta Cl$) were simulated to be
independent and normally distributed at the subject level (varying
across subjects) with means of 0 and variances of 0.01. 

A covariate “GENDER” was simulated, so that there were 50 percent
males and 50 percent females. A covariate “BODYWEIGHT” was simulated
with a mean of 70 kg for males, 65 kg for females and a standard
deviation of 15 for both groups. A covariate “Age” was simulated,
with a mean of 40 years and a standard deviation of 10. None of the
covariates had any association with the parameters, so the true
underlying model had no covariate effects.

\begin{figure}[here]
\includegraphics[scale=1]{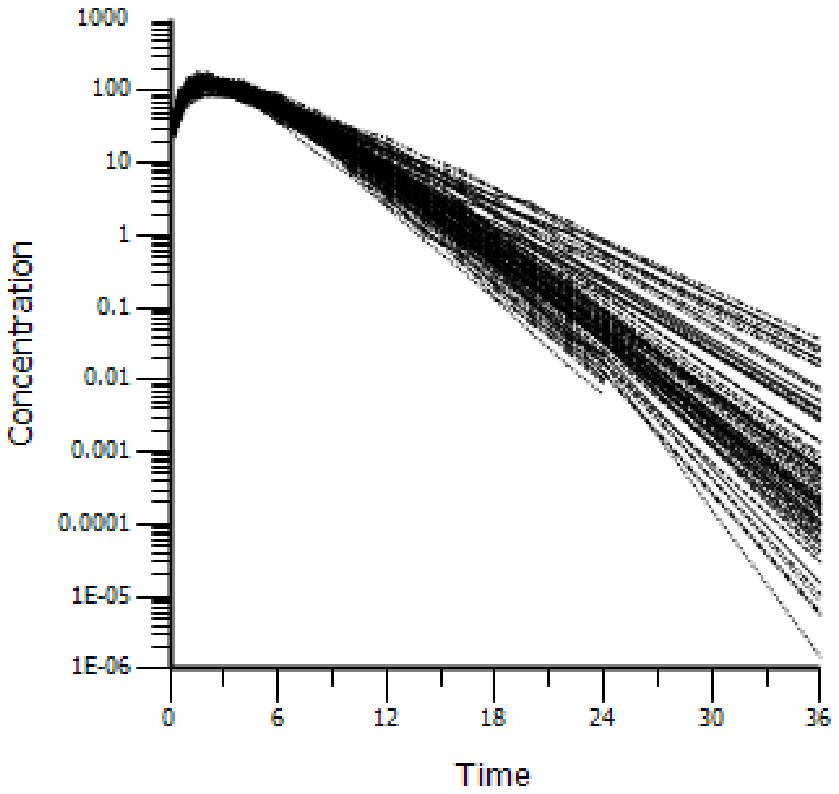}
\caption{Data simulated from a one compartment model with no covariate
  effects}
\label{F:Sim1}
\end{figure}

A dose of 5617 was administered at time 0 as an extravascular
dose. Two hundred replicates were simulated. See Figure~\ref{F:Sim1}
for a plot of the simulated data.

Pharsight’s Phoenix NLME software (version 1.3) was used to fit two
models to the simulated data. The base model was a one compartment
extravascular model with random effects for $V$ and $Cl$ and no age
effect on clearance. The full model was a one compartment
extravascular model with random effects for $V$ and $Cl$ and an age
effect on clearance. The models are specified below.

Base (correct) model:
\[
\begin{array}{rl}
    Ka &= tvKa \\
    V &= tvV \cdot \exp(\eta V) \\
    Cl &= tvCl \cdot \exp(\eta Cl) \\
\end{array}
\]

Full (incorrect) model
\[
\begin{array}{rl}
    Ka &= tvKa \\
    V &= tvV \cdot \exp(\eta V) \\
    Cl &= tvCl \cdot (\text{Age}/40)^{dCld\text{Age}} \cdot \exp(\eta
    Cl) \\
\end{array}
\]
where $tvKa$, $tvV$, $tvCl$, and $dCld\text{Age}$ are fixed effect
parameters to be estimated. Initial estimates for the fixed effect PK
parameters ($tvKa$, $tvV$, and $tvCl$) were set to the true
(simulated) parameter values. The initial estimate for the covariate
effect ($dCld\text{Age}$) was set to -3. The initial estimates of the
variances of the random effects were all 0.1, close to the true value
of 0.01.

\subsection{Simulation Example 2: Covariate Effect}
A one-compartment, extravascular model was simulated with eight
subjects using Pharsight's Trial Simulator software (version
2.2.1). The equations for the model are as follows:
\begin{align*}
\frac{dAa}{dt} &= - Ka \cdot Aa \\
\frac{dA1}{dt} &= Ka \cdot Aa - Cl \cdot C \\
C &= \frac{A1}{V} \\
\end{align*}
A 10 percent constant CV percentage was simulated for the residual
error:
\[
\text{CObs} = C(1 + C_\epsilon)
\]
where $\Var(C_\epsilon)=0.01$.
A fixed effect was added to the absorption rate parameter, $Ka$. All
other parameters were simulated with fixed and random effects. The
systemic clearance was simulated with an age effect.
\[
\begin{array}{rl}
    Ka &= tvKa \\
    V &= tvV \cdot \exp(\eta V) \\
    Cl &= tvCl \cdot (\text{Age}/40)^{dCld\text{Age}} \cdot \exp(\eta
    Cl) \\
\end{array}
\]
The fixed effects ($tvKa$, $tvV$, $tvCl$, and $dCld\text{Age}$) were
assumed to be normally distributed at the study level (varying across
replicates) with means listed below and standard deviations of 0.05,
0.1, 0.05, and 0.04, respectively.
\[
\begin{array}{rl}
    \mean(tvKa) &= 0.35 \\
    \mean(tvV) &= 13.5 \\
    \mean(tvCl) &= 1.2 \\
    \mean(dCld\text{Age}) &= -0.9 \\
\end{array}
\]
The random effects ($\eta V$ and $\eta Cl$) were simulated to be
independent and normally distributed at the subject level (varying
across subjects) with means of 0 and variances of 0.01. 

A covariate “GENDER” was simulated, so that there were 50 percent
males and 50 percent females. A covariate “BODYWEIGHT” was simulated
with a mean of 70 kg for males, 65 kg for females and a standard
deviation of 15 for both groups. A covariate “Age” was simulated,
with a mean of 40 years and a standard deviation of 10. The true
underlying model had a covariate effect, namely an age effect on
clearance (see below)

\begin{figure}[here]
\includegraphics[scale=0.7]{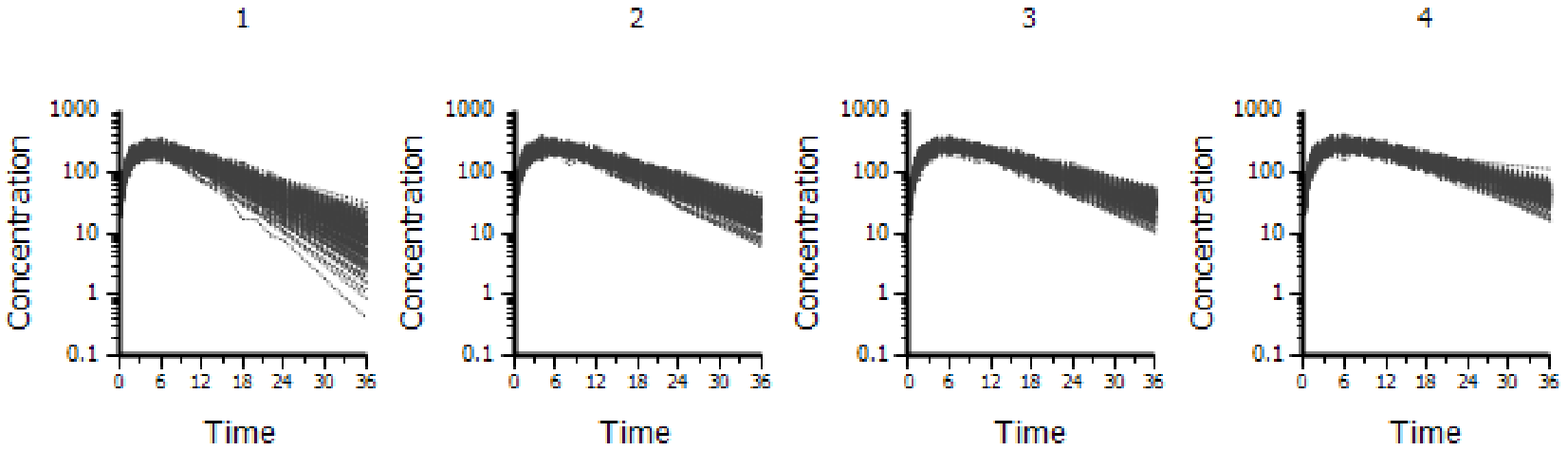}
\caption{Data simulated from one a compartment model with age effect
  on clearance, by quartiles of age}
\label{F:Sim2}
\end{figure}

A dose of 5617 was administered at time 0 as an extravascular
dose. Two hundred replicates were simulated. See Figure \ref{F:Sim2}
for a plot of the simulated data, with clearance decreasing with age.

Pharsight’s Phoenix NLME software (version 1.3) was used to fit two
models to the simulated data. The base model was a one compartment
extravascular model with random effects on $V$ and $Cl$ and no age
effect on clearance. The full model was similar to the base model, but
with an age effect included for $Cl$. The models are specified below.

Base (incorrect) model:
\[
\begin{array}{rl}
    Ka &= tvKa \\
    V &= tvV \cdot \exp(\eta V) \\
    Cl &= tvCl \cdot \exp(\eta Cl) \\
\end{array}
\]
Full (correct) model
\[
\begin{array}{rl}
    Ka &= tvKa \\
    V &= tvV \cdot \exp(\eta V) \\
    Cl &= tvCl \cdot (\text{Age}/40)^{dCld\text{Age}} \cdot \exp(\eta
    Cl) \\
\end{array}
\]
where $tvKa$, $tvV$, $tvCl$, and $dCld\text{Age}$ are fixed effects
parameters to be estimated. Initial estimates for the fixed effects
parameters ($tvKa$, $tvV$, $tvCl$, and $dCld\text{Age}$) were set to
the true (simulated) parameter values. The initial estimates of the
variances of the random effects were all 0.1, close to the true values
of 0.01.

\subsection{Simulation Example 3: Covariate Effect in Two
  Compartment Model}
A two-compartment, extravascular model was simulated with eight
subjects using Pharsight's Trial Simulator software (version
2.2.1). The equations for the model are as follows:
\begin{align*}
 \frac{dAa}{dt} &= - Ka \cdot Aa \\
 \frac{dA1}{dt} &= Ka \cdot Aa - Cl \cdot C  - Cl2 \cdot (C - C2)\\
 \frac{dA2}{dt} &= Cl2 \cdot (C - C2) \\
 C &= \frac{A1}{V} \\
 C2 &= \frac{A2}{V2} \\ 
\end{align*}
A 10 percent constant CV percentage was simulated for the residual
error:
\[
\text{CObs} = C(1 + C_\epsilon)
\]
where $\Var(C_\epsilon)=0.01$.

A fixed effect was added to the absorption rate parameter, $Ka$. All
other parameters were simulated with fixed and random effects. The
systemic clearance was simulated with an age effect.
\[
\begin{array}{rl}
    Ka &= tvKa \\
    V &= tvV \cdot \exp(\eta V) \\
    V2 &= tvV2 \cdot \exp(\eta V2) \\
    Cl &= tvCl \cdot (\text{Age}/40)^{dCld\text{Age}} \cdot \exp(\eta
    Cl) \\ 
    Cl2 &= tvCl2 \cdot \exp(\eta Cl2) \\
\end{array}
\]
The fixed effects ($tvKa$, $tvV$, $tvV2$, $tvCl$, $tvCl2$, and
$dCld\text{Age}$) were assumed to be normally distributed at the study
level (varying across replicates) with means listed below and standard
deviations of 0.05, 0.1, 0.1, 0.05, 0.05, and 0.04, respectively.
\[
\begin{array}{rl}
    \mean(tvKa) &= 0.35 \\
    \mean(tvV) &= 13.5 \\
    \mean(tvV2) &= 36 \\
    \mean(tvCl) &= 1.2 \\
    \mean(tvCl2) &= 0.62 \\
    \mean(dCld\text{Age}) &= -0.9 \\
\end{array}
\]
The random effects ($\eta V$, $\eta V2$, $\eta Cl$, and $\eta Cl2$)
were simulated to be independent and normally distributed at the
subject level (varying across subjects) with means of 0 and variances
of 0.01. 

A covariate “GENDER” was simulated, so that there were 50
percent males and 50 percent females. A covariate “BODYWEIGHT” was
simulated with a mean of 70 kg for males, 65 kg for females and a
standard deviation of 15 for both groups. A covariate “Age” was
simulated, with a mean of 40 years and a standard deviation of 10. The
true underlying model had a covariate effect, namely an age effect on
clearance (see below).

\begin{figure}[here]
\includegraphics[scale=0.7]{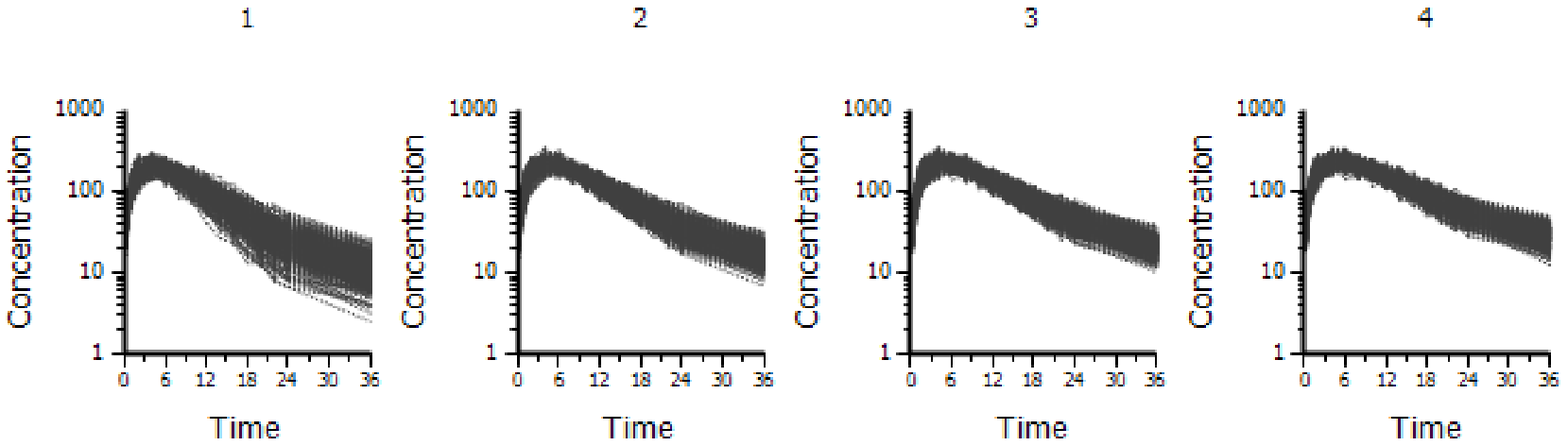}
\caption{Data simulated from a two compartment model with age effect
  on clearance, by quartiles of age}
\label{F:Sim3}
\end{figure}

A dose of 5617 was administered at time 0 as an extravascular
dose. Two hundred replicates were simulated. See Figure \ref{F:Sim3}
for a plot of the simulated data, with clearance decreasing with age.

Pharsight’s Phoenix NLME software (version 1.3) was used to fit two
models to the simulated data. The base model was a two compartment
extravascular model with random effects on $V$, $V2$, $Cl$, and $Cl2$
and no age effect on clearance. The full model was similar to the base
model, but with an age effect included for $Cl$. The models are
specified below.

Base (incorrect) model:
\[
\begin{array}{rl}
    Ka &= tvKa \\
    V &= tvV \cdot \exp(\eta V) \\
    V2 &= tvV2 \cdot \exp(\eta V2) \\
    Cl &= tvCl \cdot \exp(\eta Cl) \\
    Cl2 &= tvCl2 \cdot \exp(\eta Cl2) \\
\end{array}
\]
Full (correct) model:
\[
\begin{array}{rl}
        Ka &= tvKa \\
        V &= tvV \cdot \exp(\eta V) \\
        V2 &= tvV2 \cdot \exp(\eta V2) \\
        Cl &= tvCl \cdot (\text{Age}/40)^{dCld\text{Age}} \cdot
        \exp(\eta Cl) \\
        Cl2 &= tvCl2 \cdot \exp(\eta Cl2) \\
\end{array}
\]
where $tvKa$, $tvV$, $tvV2$, $tvCl$, $tvCl2$, and $dCld\text{Age}$ are
fixed effects parameters to be estimated. Initial estimates for the
fixed effects parameters ($tvKa$, $tvV$, $tvV2$, $tvCl$, $tvCl2$, and
$dCld\text{Age}$) were set to the true (simulated) parameter
values. The initial estimates of the variances of the random effects
were all 0.1, close to the true values of 0.01.

\subsection{Simulation Example 4: Five Covariate Effects}
A one compartment, extravascular model was simulated with eight
subjects using Pharsight's Trial Simulator software (version
2.2.1). The equations for the model are as follows:
\begin{align*}
\frac{dAa}{dt} &= - Ka \cdot Aa \\
\frac{dA1}{dt} &= Ka \cdot Aa - Cl \cdot C \\
C &= \frac{A1}{V} \\
\end{align*}
A 10 percent constant CV percentage was simulated for the residual
error:
\[
\text{CObs} = C(1 + C_\epsilon)
\]
where $\Var(C_\epsilon)=0.01$.

A fixed effect was added to the absorption rate parameter, $Ka$. All
other parameters were simulated with fixed and random effects. The
systemic volume was simulated with a body weight effect. The systemic
clearance was simulated with body weight (BW), age (Age), gender
(Gender), and hepatic impairment (HI) effects:
\[
\begin{array}{rl}
    Ka &= tvKa \\
    V &= tvV \cdot (\text{BW}/70)^{dVd\text{BW}} \cdot \exp(\eta V) \\
    Cl &= tvCl \cdot (\text{BW}/70)^{dCld\text{BW}} \cdot
    (\text{Age}/40)^{dCld\text{Age}} \cdot (1 + dCld\text{Gender}) \\
    & \cdot (1 + dCld\text{HI}*\text{HI}) \cdot \exp(\eta Cl) \\
\end{array}
\]
The fixed effects ($tvKa$, $tvV$, $tvCl$, $dVd\text{BW}$,
$dCld\text{BW}$, $dCld\text{Age}$, $dCld\text{Gender}$, and
$dCld\text{HI}$) were assumed to be normally distributed at the study
level (varying across replicates) with means listed below and standard
deviations of 0.05, 0.1, 0.05, 0.1, 0.1, 0.04, 0.05, and 0.05
respectively.
\[
\begin{array}{rl}
    \mean(tvKa) &= 0.35 \\
    \mean(tvV) &= 13.5 \\
    \mean(tvCl) &= 1.2 \\
    \mean(dVd\text{BW}) &= 1 \\
    \mean(dCld\text{BW}) &= 0.75 \\
    \mean(dCld\text{Age}) &= -0.9 \\
    \mean(dCld\text{Gender}) &= 0.1 \\
    \mean(dCld\text{HI}) &= -0.2 \\
\end{array}
\]
The random effects ($\eta V$ and $\eta Cl$) were simulated to be
independent and normally distributed at the subject level (varying
across subjects) with means of 0 and variances of 0.01. 

A covariate ``Gender'' was simulated, so that there were 50 percent
males (Gender=1) and 50 percent females (Gender=0). A covariate for
body weight ``BW'' was simulated with a mean of 70 kg for
males, 65 kg for females and a standard deviation of 15 for both
groups. A covariate ``Age'' was simulated, with a mean of 40 years and
a standard deviation of 10. A covariate for hepatic impairment ``HI''
was simulated, with 70 percent not hepatically impaired (HI=0) and 30
percent hepatically impaired (HI=1). The true underlying model had
five covariate effects, namely a body weight effect on volume, and
age, body weight, gender, and hepatic impairment effects on clearance
(see below).

\begin{figure}[here]
\includegraphics[scale=1]{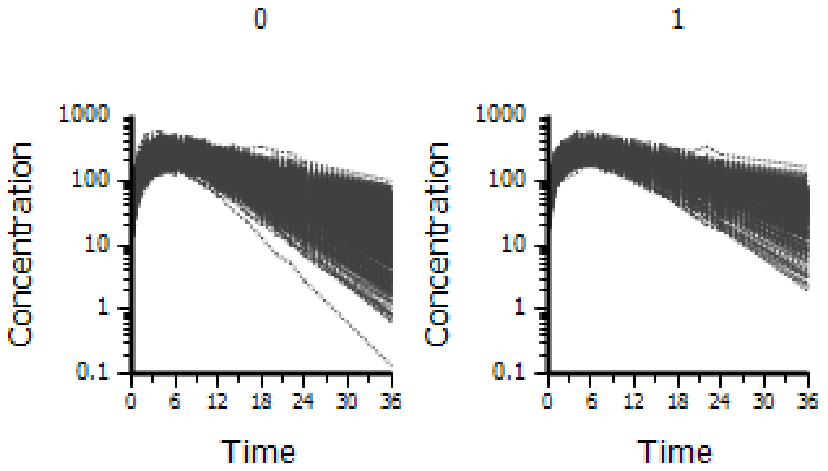}
\caption{Data simulated from a one compartment model with body weight
  effect on volume, and body weight, gender, age, and hepatic
  impairment effects on clearance, by hepatic impairment (0=Not
  hepatically impaired, 1=Hepatically impaired)}
\label{F:Sim5}
\end{figure}

A dose of 5617 was administered at time 0 as an extravascular
dose. Two hundred replicates were simulated. See Figure \ref{F:Sim5}
for a plot of the simulated data.

Pharsight’s Phoenix NLME software (version 1.3) was used to fit two
models to the simulated data. The base model was a one compartment
extravascular model with random effects on $V$ and $Cl$, a body weight
effect on $V$, and age, gender, and body weight effects on $Cl$. The
full model was similar to the base model, but with a hepatic
impairment effect also included for $Cl$. The models are specified
below.

Base (incorrect) model:
\[
\begin{array}{rl}
    Ka &= tvKa \\
    V &= tvV \cdot (\text{BW}/70)^{dVd\text{BW}} \cdot \exp(\eta V) \\
    Cl &= tvCl \cdot (\text{BW}/70)^{dCld\text{BW}} \cdot
    (\text{Age}/40)^{dCld\text{Age}} \cdot (1 + dCld\text{Gender})
    \cdot \exp(\eta Cl) \\
\end{array}
\]
Full (correct) model:
\[
\begin{array}{rl}
    Ka &= tvKa \\
    V &= tvV \cdot (\text{BW}/70)^{dVd\text{BW}} \cdot \exp(\eta V) \\
    Cl &= tvCl \cdot (\text{BW}/70)^{dCld\text{BW}} \cdot
    (\text{Age}/40)^{dCld\text{Age}} \cdot (1 + dCld\text{Gender}) \\
    & \cdot (1 + dCld\text{HI}*\text{HI}) \cdot \exp(\eta Cl) \\
\end{array}
\]
where $tvKa$, $tvV$, $tvCl$, $dVd\text{BW}$, $dCld\text{BW}$,
$dCld\text{Age}$, $dCld\text{Gender}$, and $dCld\text{HI}$ are fixed
effects parameters to be estimated. Initial estimates for the fixed
effects parameters were set to the true (simulated) parameter
values. The initial estimates of the variances of the random effects
were all 0.1, close to the true values of 0.01.

\subsection{Simulation Example 5: Two compartment model}
A two compartment, extravascular model was simulated with six subjects
using Pharsight's Trial Simulator software (version 2.2.1). The
equations for the model are as follows:
\begin{align*}
 \frac{dAa}{dt} &= - Ka \cdot Aa \\
 \frac{dA1}{dt} &= Ka \cdot Aa - Cl \cdot C  - Cl2 \cdot (C - C2)\\
 \frac{dA2}{dt} &= Cl2 \cdot (C - C2) \\
 C &= \frac{A1}{V} \\
 C2 &= \frac{A2}{V2} \\ 
\end{align*}
A 10 percent constant CV percentage was simulated for the residual
error:
\[
\text{CObs} = C(1 + C_\epsilon)
\]
where $\Var(C_\epsilon)=0.01$.

A fixed effect was added to the absorption rate parameter, $Ka$. All
other parameters were simulated with fixed and random effects:
\[
\begin{array}{rl}
    Ka &= tvKa \\
    V &= tvV \cdot \exp(\eta V) \\
    V2 &= tvV2 \cdot \exp(\eta V2) \\
    Cl &= tvCl \cdot \exp(\eta Cl) \\
    Cl2 &= tvCl2 \cdot \exp(\eta Cl2) \\
\end{array}
\]
The fixed effects for the PK parameters were assumed to be normally
distributed at the study level (varying across replicates) with means
listed below and standard deviations of 0.1, except in this example
the fixed effect for $Ka$ was simulated with a standard deviation of
0.05 because when the absorption rate was smaller the portion of the
curve for the first compartment became less pronounced in relation to
the portion for the second compartment. Having a smaller standard
deviation for $Ka$ increased the chance that all the simulated
profiles would have a characteristic two compartment shape.
\[
\begin{array}{rl}
    \mean(tvKa) &= 0.35 \\
    \mean(tvV) &= 13.5 \\
    \mean(tvV2) &= 34 \\
    \mean(tvCl) &= 7.4 \\
    \mean(tvCl2) &= 1.2 \\
\end{array}
\]
The random effects ($\eta V$, $\eta V2$, $\eta Cl$, and $\eta Cl2$)
were simulated to be normally distributed at the subject level
(varying across subjects) with means of 0 and variances of 0.01. 

\begin{figure}[here]
\includegraphics[scale=1]{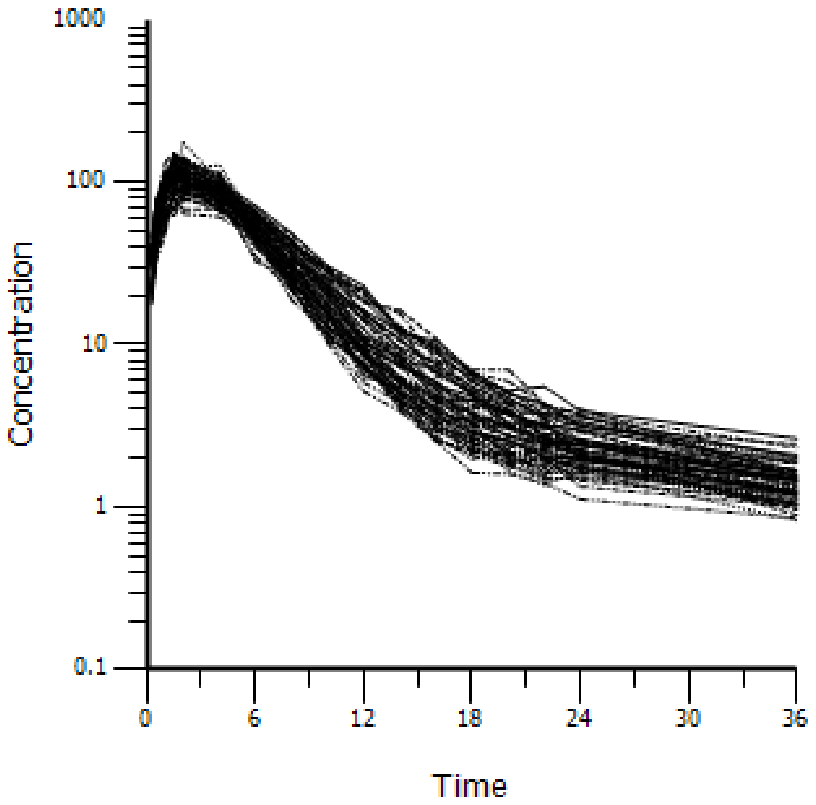}
\caption{Data simulated from a two compartment model with no covariate
  effects}
\label{F:Sim6}
\end{figure}

A dose of 5617 was administered at time 0 as an extravascular
dose. One hundred replicates were simulated. See Figure \ref{F:Sim6}
for a plot of the simulated data.

Pharsight’s Phoenix NLME software (version 1.3) was used to fit a base
model with one compartment and a full model with two compartments to
the simulated data. The models are specified below.

Base (incorrect) model
\[
\begin{array}{rl}
    Ka &= tvKa \\
    V &= tvV \cdot \exp(\eta V) \\
    Cl &= tvCl \cdot \exp(\eta Cl) \\
\end{array}
\]

Full (correct) model
\[
\begin{array}{rl}
    Ka &= tvKa \\
    V &= tvV \cdot \exp(\eta V) \\
    V2 &= tvV2 \cdot \exp(\eta V2) \\
    Cl &= tvCl \cdot \exp(\eta Cl) \\
    Cl2 &= tvCl2 \cdot \exp(\eta Cl2) \\
\end{array}
\]
where $tvKa$, $tvV$, $tvV2$, $tvCl$, and $tvCl2$ are fixed effects
parameters to be estimated. Initial estimates for the fixed effects
parameters ($tvKa$, $tvV$, $tvV2$, $tvCl$, and $tvCl2$) were set to
the true (simulated) parameter values. The initial estimates of the
variances of the random effects were all 0.1, close to the true values
of 0.01.

\section{Distribution of the Selection Statistics and Shrinkage in the
  Simulated Data Sets} \label{SS:sim_dist}
\begin{figure}[here]
\includegraphics[scale=0.6]{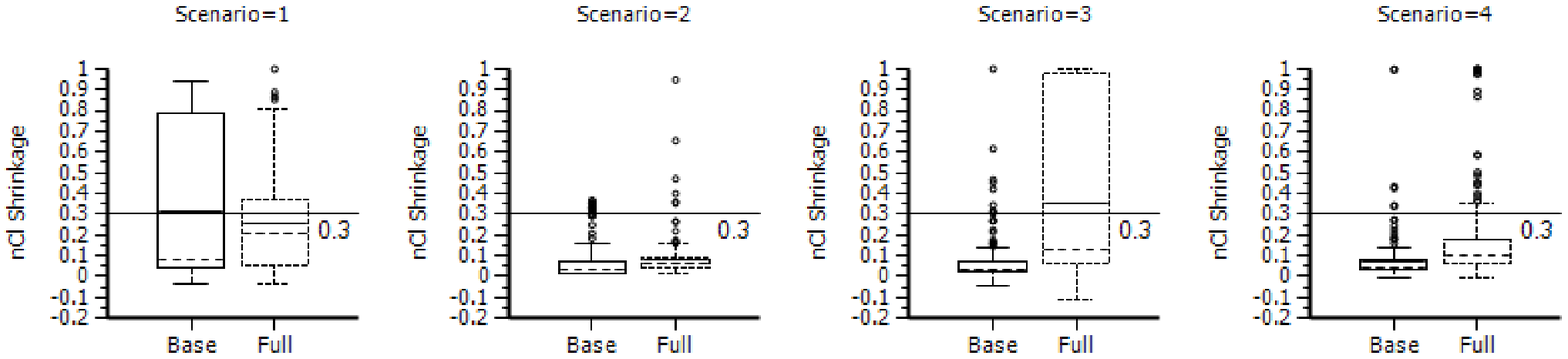}
\caption{Boxplots of the $\eta$ shrinkage values for each model under
  scenarios 1-4}
\label{F:shrinkage}
\end{figure}

For each simulated data set, the values of AIC, BIC,
$\text{CrV}_\eta$, $\text{CrV}_y$, and $\text{wtCrV}_y$ were
calculated, along with the standard error of each statistic. Tables
\ref{T:sim_AICBIC}, \ref{T:sim_npress}, \ref{T:sim_mpress}, and
\ref{T:sim_wtmpress} show the mean values and standard deviations of
both the statistics and the standard errors over 200 simulations (or
100 simulations in the case of scenario 5). Furthermore, boxplots of
the $\eta$ shrinkage values for each model under scenarios 1-4 are
shown in Figure \ref{F:shrinkage}.

\begin{table}
\caption{Distribution of AIC and BIC in each simulation scenario}
\label{T:sim_AICBIC}
\begin{center}
\begin{tabular}{cccccccc}
\hline\noalign{\smallskip}
& & & & \multicolumn{2}{c}{AIC}& \multicolumn{2}{c}{BIC}  \\	
Scen.& Base Model  &        Full Model& & Base& Full& Base& Full \\
\noalign{\smallskip}\hline\noalign{\smallskip}
1& \checkmark 1 Cpt&   	 1 Cpt,& N& 200& 200& 200& 200 \\
&     	    & 		     Age-Cl& Mean& 384& 404& 401& 424 \\
&			&			       & SD& 114& 115& 114& 115 \\
2& 1 Cpt	& \checkmark 1 Cpt,& N& 200& 200& 200& 200 \\
&    		& 	        Age-Cl & Mean& 1106& 1093& 1124& 1113 \\
&			&	               & SD& 25& 25& 25& 25 \\
3& 2 Cpt	& \checkmark 2 Cpt,& N& 200& 200& 200& 200 \\
&    		&	       Age-Cl  & Mean& 1039& 1026& 1068& 1058 \\
&			&			  & SD& 27& 27& 29& 29 \\
4& 1 Cpt, BW-V;& \checkmark 1 Cpt, BW-V;& N& 200& 200& 200& 200 \\
& BW-Cl, G-Cl,& BW-Cl, G-Cl,& Mean& 1106& 1100& 1135& 1132 \\
& Age-Cl     &	 	Age-Cl, HI-Cl& SD& 33& 32& 33& 32 \\
5&  1 Cpt	& \checkmark 2 Cpt& N& 100& 100& 100& 100 \\
&			&			  & Mean& 648& 417& 664& 443 \\
&			&			  & SD& 218& 26& 218& 26 \\
\noalign{\smallskip}\hline
\end{tabular}
\end{center}
\checkmark indicates true model, Cpt=Compartment, Age-Cl indicates age effect on clearance, BW=Body
Weight, V=Volume, G=Gender, HI=Hepatic Impairment
\end{table}
 
\begin{table}
\caption{Distribution of $\text{CrV}_\eta$ in each simulation scenario}
\label{T:sim_npress}
\begin{center}
\begin{tabular}{cccccccc}
\hline\noalign{\smallskip}
&& & & \multicolumn{2}{c}{$\text{CrV}_\eta$}&	\multicolumn{2}{c}{SE($\text{CrV}_\eta$)} \\
Scen.& Base Model& Full Model& &	Base&	Full&	Base&	Full \\
\noalign{\smallskip}\hline\noalign{\smallskip}
1& \checkmark 1 Cpt& 	 	1 Cpt,&	N&	200&	200&	200&	200 \\
& 				   &  		Age-Cl&	Mean&	0.136&	0.919&	0.022&	0.403 \\
& 				   &	  	      &	SD&	0.236&	0.920&	0.032&	0.512 \\
2& 			 1 Cpt,& \checkmark 1 Cpt,&	N&	200&	200&	200&	200 \\
& 			 		& 	 	  Age-Cl&	Mean&	0.078&	0.012&	0.031&	0.005 \\
& 				   & 				&	SD&	0.056&	0.010&	0.022&	0.005 \\
3& 			 2 Cpt & \checkmark 2 Cpt,&	N&	200&	200&	200&	200 \\
& 		           & 		  Age-Cl&	Mean&	0.144&	0.024&	0.077&	0.020 \\
& 				   &				&	SD&	0.728&	0.155&	0.453&	0.155 \\
4& 	   1 Cpt, BW-V;& \checkmark 1 Cpt, BW-V;&	N&	200&	200&	200&	200 \\
& 		BW-Cl, G-Cl,& 	BW-Cl, G-Cl,&	Mean&	1.700&	0.124&	0.626&	0.081 \\
& 		Age-Cl&		  Age-Cl, HI-Cl&	SD&	2.577&	0.281&	2.515&	0.214 \\
\noalign{\smallskip}\hline
\end{tabular}
\end{center}
\checkmark indicates true model, Cpt=Compartment, Age-Cl indicates age effect on clearance, BW=Body
Weight, V=Volume, G=Gender, HI=Hepatic Impairment
\end{table}

\begin{table}
\caption{Distribution of $\text{CrV}_y$ in each simulation scenario}
\label{T:sim_mpress}
\begin{center}
\begin{tabular}{cccccccc}
\hline\noalign{\smallskip}
& & & & \multicolumn{2}{c}{$\text{CrV}_y$}&	\multicolumn{2}{c}{SE($\text{CrV}_y$)} \\
Scen.& Base Model& Full Model& &	Base&	Full&	Base&	Full \\
\noalign{\smallskip}\hline\noalign{\smallskip}
1& \checkmark 1 Cpt& 1 Cpt,&	N&	200&	200&	200&	200 \\
& & Age-Cl&	Mean&	41.1&	46.4&	13.3&	18.5 \\
& & &	SD&	58.4&	82.0&	51.6&	73.6 \\
2& 1 Cpt& \checkmark 1 Cpt,&	N&	200&	200&	200&	200 \\
& & Age-Cl&	Mean&	247&	254&	45.6&	 47.2 \\
& & &	SD&	54.0&	59.0&	24.4&	25.5 \\
3& 2 Cpt& \checkmark 2 Cpt,&	N&	200&	200&	200&	200 \\
& & Age-Cl&	Mean&	159&	199&	28.0&	59.9 \\
& & &	SD&	34.9&	171&	12.5&	160 \\
4& 1 Cpt, BW-V;& \checkmark 1 Cpt, BW-V;&	N&	200&	200&	200&	200 \\
& BW-Cl, G-Cl,& BW-Cl, G-Cl,&	Mean&	323&	472&	85.1&	173 \\
& Age-Cl& Age-Cl, HI-Cl&	SD&	116&	401&	72.2&	284 \\
5& 1 Cpt& \checkmark 2 Cpt&	N&	100&	100&	100&	100 \\
& & &	Mean&	206.11&	30.28&	32.32&	8.32 \\
& & &	SD&	90.3&	9.60&	20.7&	5.74 \\
\noalign{\smallskip}\hline
\end{tabular}
\end{center}
\checkmark indicates true model, Cpt=Compartment, Age-Cl indicates age effect on clearance, BW=Body
Weight, V=Volume, G=Gender, HI=Hepatic Impairment \\
\end{table}

\begin{table}
\caption{Distribution of $\text{wtCrV}_y$ in each simulation scenario}
\label{T:sim_wtmpress}
\begin{center}
\begin{tabular}{cccccccc}
\hline\noalign{\smallskip}
& & & & \multicolumn{2}{c}{$\text{wtCrV}_y$}&	\multicolumn{2}{c}{SE($\text{wtCrV}_y$)} \\
Scen.& Base Model& Full Model& &	Base&	Full&	Base&	Full \\
\noalign{\smallskip}\hline\noalign{\smallskip}
1& \checkmark 1 Cpt& 1 Cpt,&	N&	200&	200&	200&	200 \\
& & Age-Cl&	Mean&	1.14&	1.20&	0.157&	0.176 \\
& & &	SD&	0.303&	0.388&	0.064&	0.154 \\
2& 1 Cpt& \checkmark 1 Cpt,&	N&	200&	200&	200&	200 \\
& & Age-Cl&	Mean&	0.967&	1.01&	0.144&	 0.154 \\
& & &	SD&	0.133&	0.130&	0.120&	0.111 \\
3& 2 Cpt& \checkmark 2 Cpt,&	N&	200&	200&	200&	200 \\
& & Age-Cl&	Mean&	1.0096&	1550*&	0.154&	1550* \\
& & &	SD&	0.0995&	21700*&	0.071&	21700* \\
4& 1 Cpt, BW-V;& \checkmark 1 Cpt, BW-V;&	N&	200&	200&	200&	200 \\
& BW-Cl, G-Cl,& BW-Cl, G-Cl,&	Mean&	1.25&	2.78&	0.305&	1.56 \\
& Age-Cl& Age-Cl, HI-Cl&	SD&	0.485&	11.6&	0.432&	11.3 \\
5& 1 Cpt& \checkmark 2 Cpt&	N&	100&	100&	100&	100 \\
& & &	Mean&	14.7&	1.14&	3.09&	0.233 \\
& & &	SD&	10.1&	0.143&	2.86&	0.130 \\
\noalign{\smallskip}\hline
\end{tabular}
\end{center}
\checkmark indicates true model, Cpt=Compartment, Age-Cl indicates age effect on clearance, BW=Body
Weight, V=Volume, G=Gender, HI=Hepatic Impairment \\
*One of the replicates (161) had an inflated value for
$\text{wtCrV}_y$. One subject was significantly younger than the
others, and the effect of age on clearance was estimated to be around
-22 instead of the true value of -0.9 in the model where this subject
was left out. Hence the younger subject had inflated residuals.
\end{table}

\section{Additional Details for the Indomethacin Data Example}

\subsection{Model for the Indomethacin Data Analysis}
\label{SS:indometh_desc}
In this analysis, we compared the following two compartment IV bolus model
\begin{align*}
 \frac{dA1}{dt} &= - Cl \cdot C  - Cl2 \cdot (C - C2)\\
 \frac{dA2}{dt} &= Cl2 \cdot (C - C2) \\
 C &= \frac{A1}{V} \\
 C2 &= \frac{A2}{V2}
\end{align*}
to the corresponding one compartment model:
\begin{align*}
 \frac{dA1}{dt} &= - Cl \cdot C \\
 C &= \frac{A1}{V}
\end{align*}
In the above equations, we use the same notation that was defined in
Section \ref{SS:notation}. A constant CV percentage was modeled for
the residual error in both cases: 
\[
\text{CObs} = C(1 + C_\epsilon)
\]
Under the full (two compartment) model, all PK parameters were modeled
with fixed and random effects:
\[
\begin{array}{rl}
    V &= tvV \cdot \exp(\eta V) \\
    V2 &= tvV2 \cdot \exp(\eta V2) \\
    Cl &= tvCl \cdot \exp(\eta Cl) \\
    Cl2 &= tvCl2 \cdot \exp(\eta Cl2)
\end{array}
\]
where $tvKa$, $tvV$, $tvV2$, $tvCl$, and $tvCl2$ are fixed effects
parameters to be estimated. The corresponding base (one compartment)
model was similar:
\[
\begin{array}{rl}
    V &= tvV \cdot \exp(\eta V) \\
    Cl &= tvCl \cdot \exp(\eta Cl)
\end{array}
\]
The fixed effects for the PK parameters were set to the following
initial values.
\[
\begin{array}{rl}
    tvV &= 7551 \\
    tvV2 &= 13531 \\
    tvCl &= 8368 \\
    tvCl2 &= 7056 \\
\end{array}
\]
The random effects ($\eta V$, $\eta V2$, $\eta Cl$, and $\eta Cl2$)
were set to have initial variances of 1.

\subsection{Parameter Estimates and Diagnostic Plots for the
  Indomethacin Data Analysis} \label{SS:indometh_diag}
\begin{figure}[here]
\includegraphics[scale=0.4]{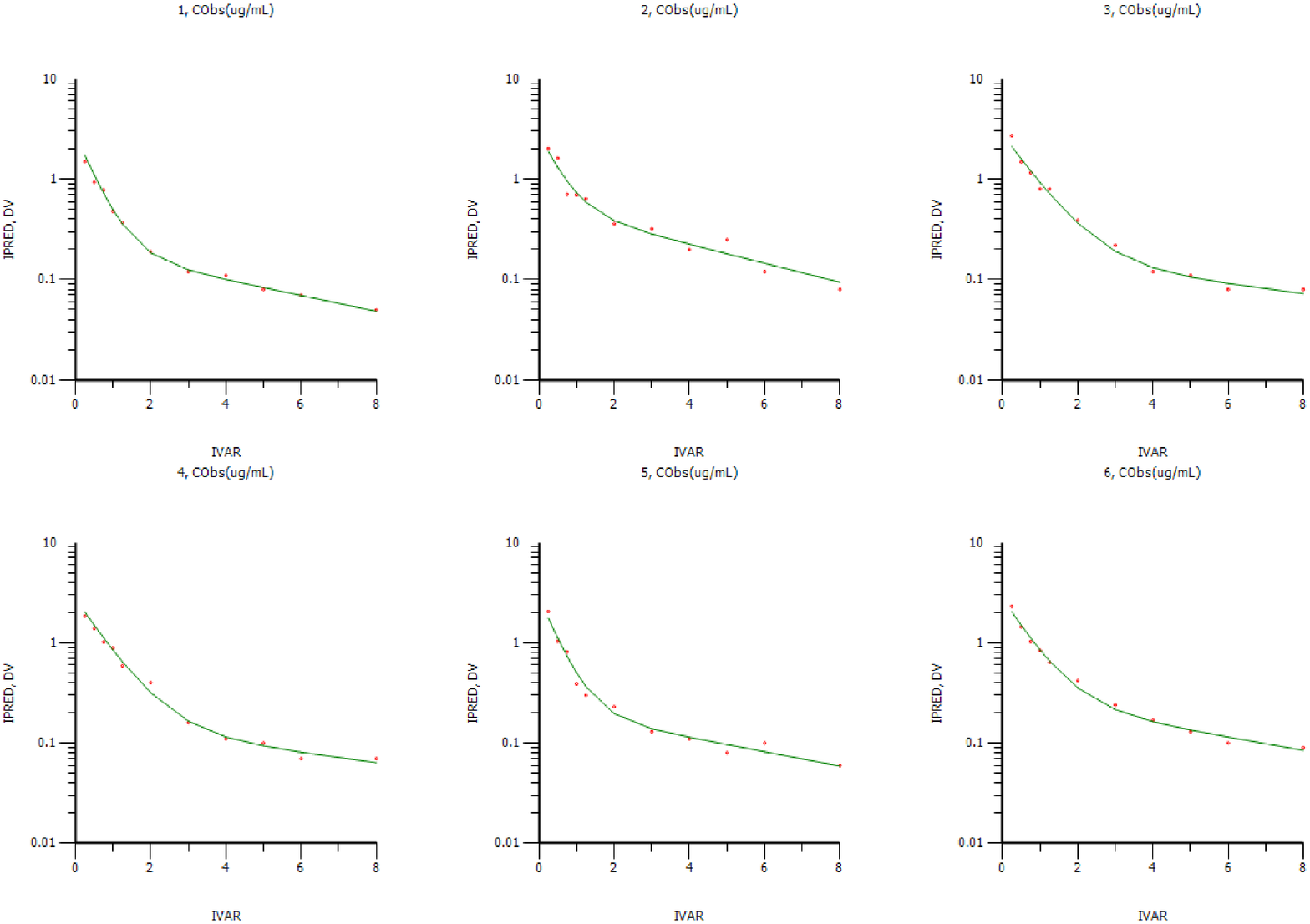}
\caption{Predicted (lines) and observed (dots) concentrations versus
  time for the indomethacin model}
\label{F:indometh2}
\end{figure}

\begin{figure}[here]
\includegraphics[scale=0.75]{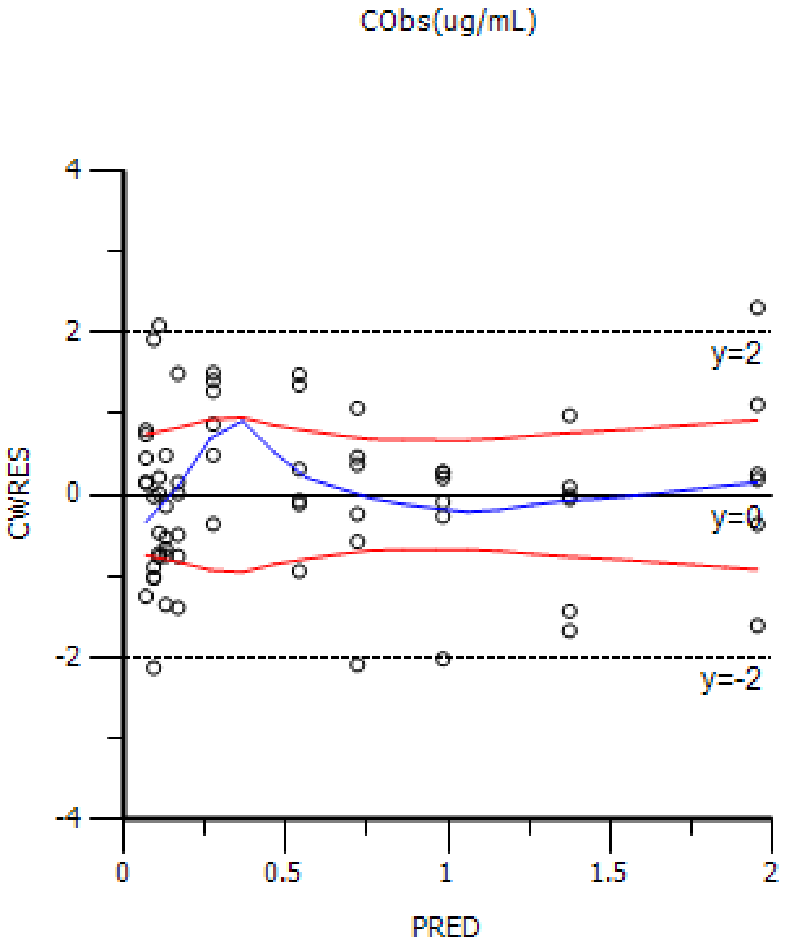}
\caption{Conditional weighted residuals versus predicted values
  for the indomethacin model}
\label{F:indometh3}
\end{figure}

\begin{figure}[here]
\includegraphics[scale=0.75]{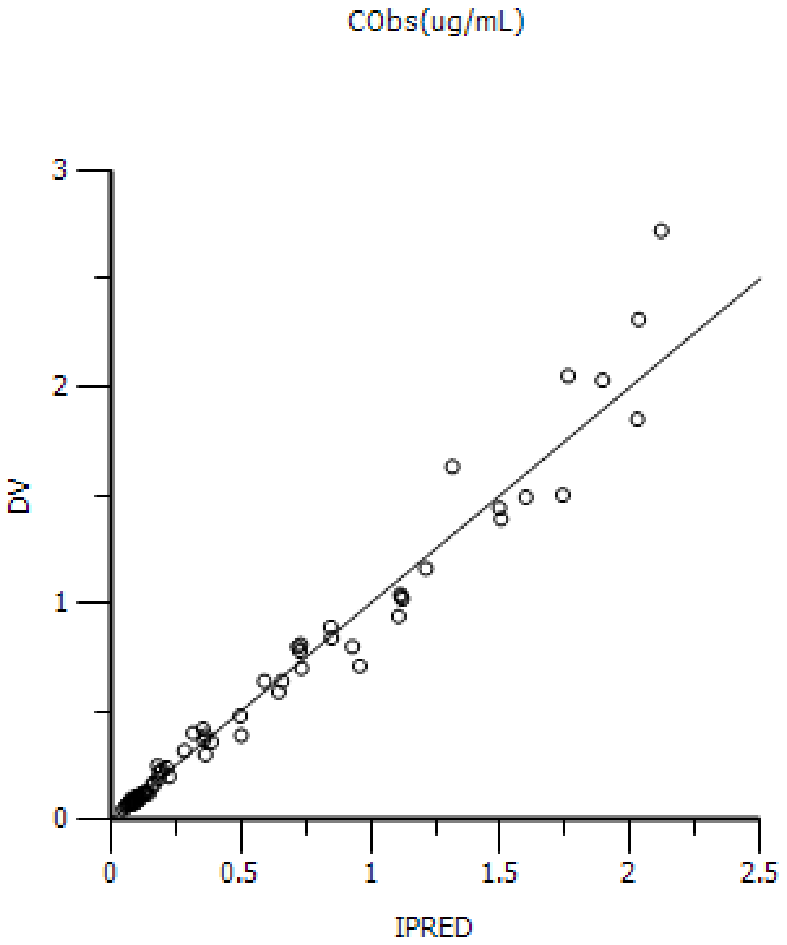}
\caption{Observed versus individual predicted values for the
  indomethacin model}
\label{F:indometh4}
\end{figure}

The coefficient estimates for the fixed effects of the indomethacin
model are shown in Table \ref{T:theta_indometh}, and the estimated
variance-covariance matrix is shown in Table \ref{T:omega_indometh}.
To assess the model fit, three diagnostic plots were created. Figure
\ref{F:indometh2} shows the predicted and observed concentrations over
time for all six subjects in the data set. Figure \ref{F:indometh3}
shows a plot of the conditional weighted residuals versus the
predicted concentrations. Figure \ref{F:indometh4} shows a scatter plot
of the observed concentrations versus the predicted concentrations. In
each case, no obvious problems are visible, suggesting that the model
fit is adequate.

\begin{table}
\caption{Parameter estimates for the indomethacin model}
\label{T:theta_indometh}
\begin{center}
\begin{tabular}{ccccc}
\hline\noalign{\smallskip}
Parameter&	Estimate&	Units&	Stderr& Stderr\% \\
\noalign{\smallskip}\hline\noalign{\smallskip}
$tvV$  &	8898&	mL&		574.84&	6.46 \\
$tvV2$ &	19527.3&	mL&	3169.70&	16.23 \\
$tvCl$ &	7905.99&	mL/h&	608.53&	7.70 \\
$tvCl2$&	5252.15&	mL/h&	768.93&	14.64 \\
$\sigma$ &	0.1440&	 &   0.02&	13.25 \\
\noalign{\smallskip}\hline
\end{tabular}
\end{center}
$\sigma$ denotes the estimated residual standard deviation\\
prefix of 'tv' denotes fixed effect or typical value
\end{table} 

\begin{table}
\caption{Estimated variance/covariance matrix ($\Omega$) of the random
  effects for the indomethacin model}
\label{T:omega_indometh}
\begin{center}
\begin{tabular}{ccccc}
\hline\noalign{\smallskip}
 &	$\eta V$&	$\eta Cl$&	$\eta V2$& $\eta Cl2$ \\
\noalign{\smallskip}\hline\noalign{\smallskip}
$\eta V$  &	0.0017&	&	&	\\
$\eta Cl$ &	0&	0.0338	& &	\\
$\eta V2$ &	0&	0&	0.0666 &	\\
$\eta Cl2$&	0&	0&	0&	0.1202 \\
Shrinkage &	0.7064&	0.0329&	0.3321&	0.0727 \\
\noalign{\smallskip}\hline
\end{tabular}
\end{center}
\end{table}

\section{Additional Details for the Theophylline Data Example}

\subsection{Model for the Theophylline Data Analysis}
\label{SS:theoph_desc}
A one compartment extravascular model with clearance parametrization
was fit to this data set. The equations for the model are as follows:
\begin{align*}
 \frac{dAa}{dt} &= - Ka \cdot Aa \\
 \frac{dA1}{dt} &= Ka \cdot Aa - Cl \cdot C \\
 C &= \frac{A1}{V} \\
\end{align*}
In the above equations, we use the same notation that was defined in
Section \ref{SS:notation}. The dose to the absorption compartment is
delayed by a parameter Tlag. An additive weighting scheme was modeled
for the residual error:
\[
\text{CObs} = C + C_\epsilon
\]
A fixed effect was added to the time lag parameter, Tlag. All
other parameters were modeled with fixed and random effects:
\[
\begin{array}{rl}
    Ka &= tvKa \cdot (\text{wt}/\mean(\text{wt}))^{dKad\text{wt}}
    \cdot \exp(\eta Ka) \\
    V &= tvV \cdot \exp(\eta V) \\
    Cl &= tvCl \cdot \exp(\eta Cl) \\
    \text{Tlag} &= tv\text{Tlag} \\
\end{array}
\]
where $tvKa$, $tvV$, $tvV2$, $tvCl$, and $tvCl2$ are fixed effects
parameters to be estimated, and wt is the body weight of a subject
in kg and $\mean(\text{wt})$ is the average of the body weights in the
data set. This full model was compared to two base models. The first
base model did not include a Tlag parameter nor a covariate for
weight:
\[
\begin{array}{rl}
    Ka &= tvKa \cdot \exp(\eta Ka) \\
    V &= tvV \cdot \exp(\eta V) \\
    Cl &= tvCl \cdot \exp(\eta Cl)
\end{array}
\]
The second base model included a Tlag parameter but no covariate for
weight:
\[
\begin{array}{rl}
    Ka &= tvKa \cdot \exp(\eta Ka) \\
    V &= tvV \cdot \exp(\eta V) \\
    Cl &= tvCl \cdot \exp(\eta Cl) \\
    \text{Tlag} &= tv\text{Tlag} \\
\end{array}
\]
The full model included a Tlag parameter and a covariate for
weight:
\[
\begin{array}{rl}
    Ka &= tvKa \cdot (\text{wt}/mean(\text{wt})^{dKad\text{wt}} \cdot \exp(\eta Ka) \\
    V &= tvV \cdot \exp(\eta V) \\
    Cl &= tvCl \cdot \exp(\eta Cl) \\
    \text{Tlag} &= tv\text{Tlag} \\
\end{array}
\]
The fixed effects for the PK parameters were set to the following
initial values:
\[
\begin{array}{rl}
    tvKa &= 2 \\
    tvV &= 1 \\
    tvCl &= 1 \\
    tvTlag &= 1 \\
    dKad\text{wt} &= 0 \\
\end{array}
\]
The random effects ($\eta V$, $\eta Ka$, and $\eta Cl$) were set to
have initial variances of 1 in a diagonal variance-covariance matrix.

\subsection{Parameter Estimates and Diagnostic Plots for the
  Theophylline Data Analysis} \label{SS:theoph_diag}
\begin{figure}[here]
\includegraphics[scale=0.4]{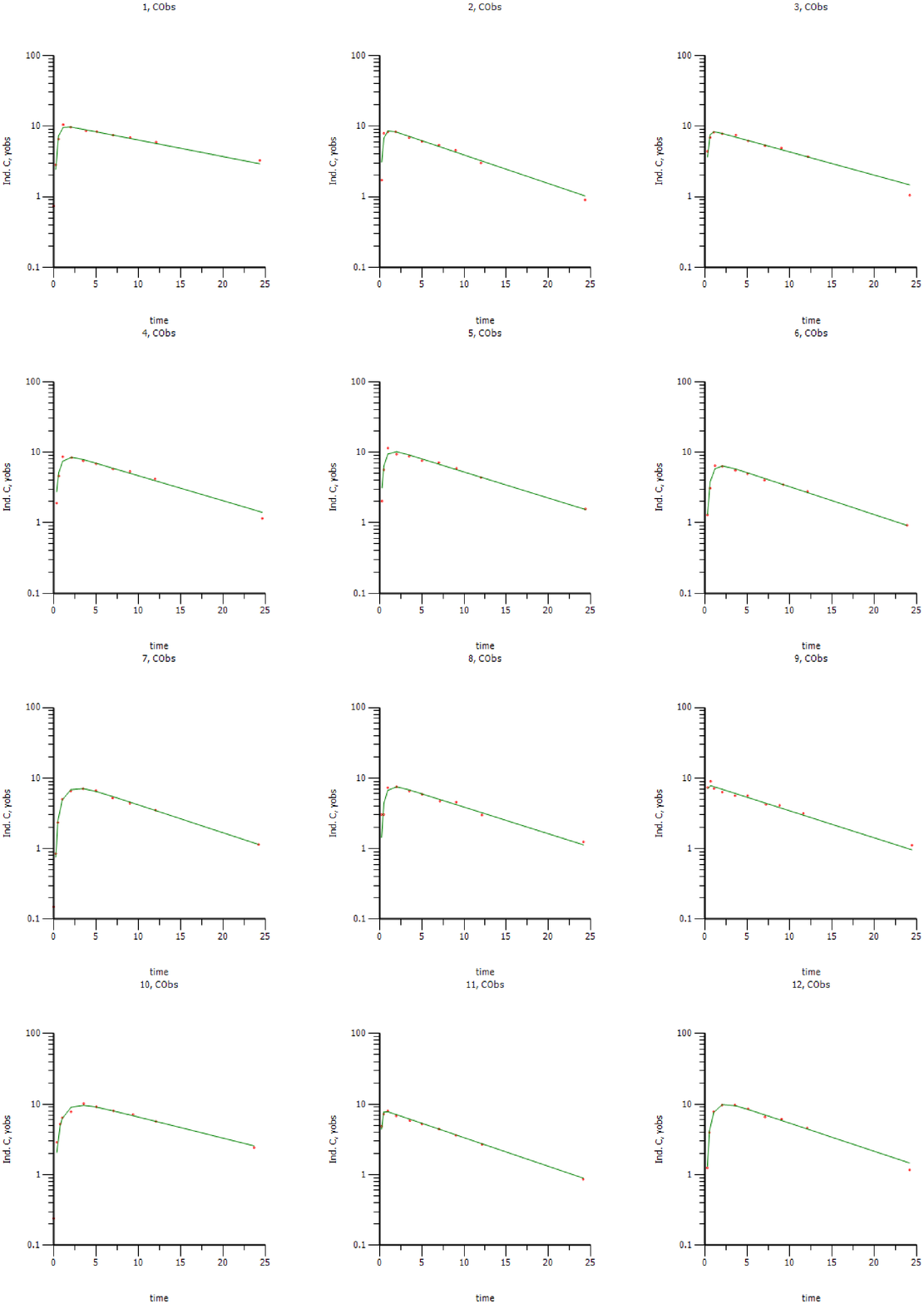}
\caption{Predicted (lines) and observed (dots) concentrations versus
  time for the final theophylline model}
\label{F:theoph2a}
\end{figure}

\begin{figure}[here]
\includegraphics[scale=0.75]{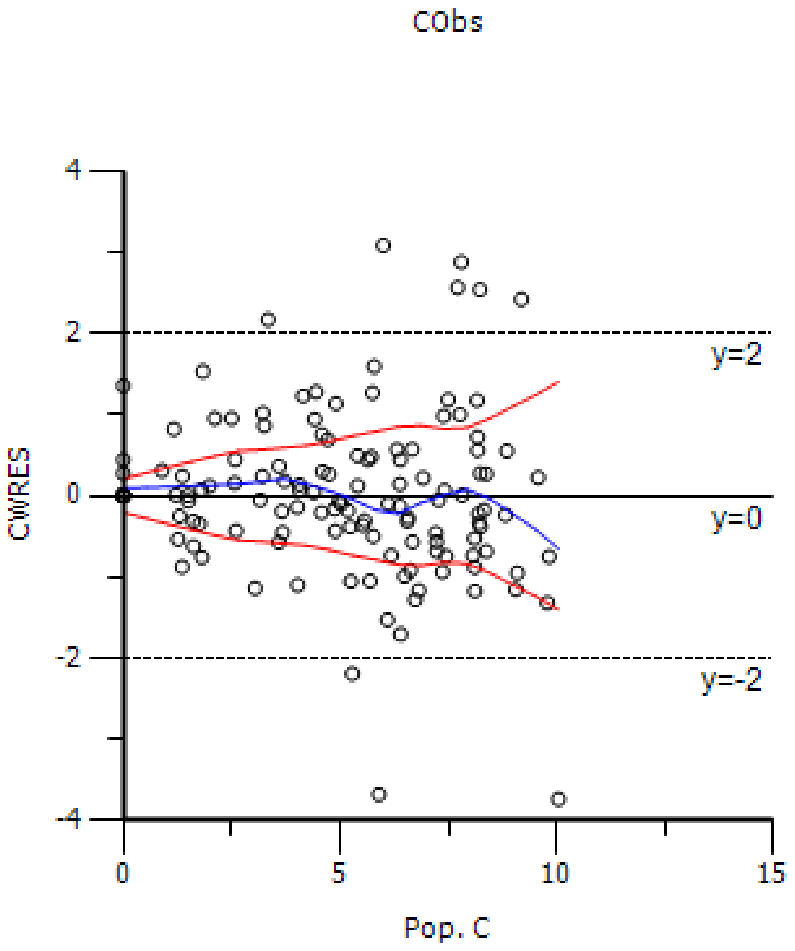}
\caption{Conditional weighted residuals versus predicted values
  for the final theophylline model}
\label{F:theoph3}
\end{figure}

\begin{figure}[here]
\includegraphics[scale=0.75]{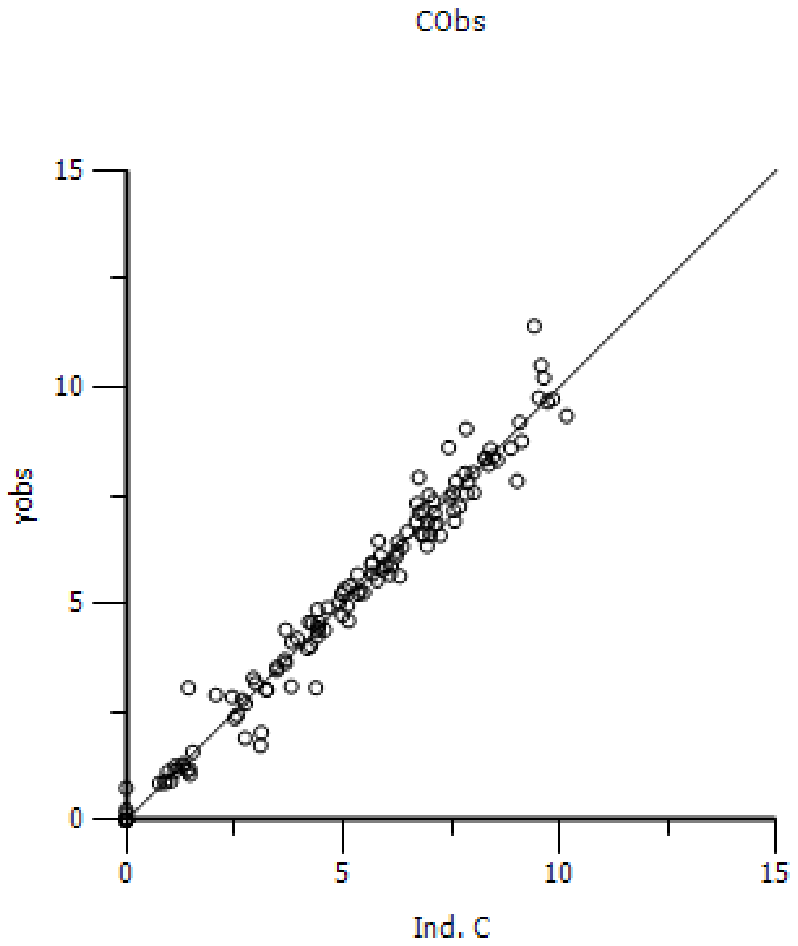}
\caption{Observed versus individual predicted values for the
  final theophylline model}
\label{F:theoph4}
\end{figure}

The coefficient estimates for the fixed effects of the indomethacin
model are shown in Table \ref{T:theta_theoph}, and the estimated
variance-covariance matrix is shown in Table \ref{T:omega_theoph}.
To assess the model fit, three diagnostic plots were created. Figure
\ref{F:theoph2a} shows the predicted and observed concentrations over
time for all six subjects in the data set. Figure \ref{F:theoph3}
shows a plot of the conditional weighted residuals versus the
predicted concentrations. Figure \ref{F:theoph4} shows a scatter plot
of the observed concentrations versus the predicted concentrations. In
each case, no obvious problems are visible, suggesting that the model
fit is adequate.

\begin{table}
\caption{Parameter estimates for the final theophylline model}
\label{T:theta_theoph}
\begin{center}
\begin{tabular}{cccc}
\hline\noalign{\smallskip}
Parameter&	Estimate&	Stderr& Stderr\% \\
\noalign{\smallskip}\hline\noalign{\smallskip}
$tvKa$	&	2.63	&	0.37	&	13.89	\\
$tvV$	&	0.47	&	0.02	&	4.36	\\
$tvCl$	&	0.04	&	0.00	&	8.01	\\
$tvTlag$	&	0.16	&	0.01	&	9.18	\\
$tvdKadwt$	&	3.07	&	0.76	&	24.81	\\
$\sigma$	&	0.55	&	0.04	&	7.16	\\
\noalign{\smallskip}\hline
\end{tabular}
\end{center}
$\sigma$ denotes the estimated residual standard deviation\\
prefix of 'tv' denotes fixed effect or typical value
\end{table} 

\begin{table}
\caption{Estimated variance/covariance matrix ($\Omega$) of the random
  effects for the final theophylline model}
\label{T:omega_theoph}
\begin{center}
\begin{tabular}{cccc}
\hline\noalign{\smallskip}
 &	$\eta V$&	$\eta Ka$& $\eta Cl$ \\
\noalign{\smallskip}\hline\noalign{\smallskip}
$\eta V$  &	0.0196&	&		\\
$\eta Ka$ &	0&	0.5208&  \\
$\eta Cl$&	0&	0&	0.0698 \\
Shrinkage &	0.0728&	0.0276&	0.0552 \\
\noalign{\smallskip}\hline
\end{tabular}
\end{center}
\end{table}

\end{document}